\documentclass[aps,superscriptaddress,showpacs,showkeys,nofootinbib,floatfix]{revtex4}
\usepackage{graphicx,epsfig,wrapfig,amssymb}
\usepackage{color}
\usepackage{amsfonts}
\usepackage{amsmath}
\usepackage{latexsym}      
\usepackage[english]{babel}
\usepackage{slashed}
\usepackage{amsthm}
\usepackage{amsfonts}
\usepackage{amssymb}
\usepackage{inputenc}
\usepackage{epsfig}
\usepackage{graphics}
\definecolor{darkgreen}{rgb}{0,0.65,0}

\newcommand{\be}{\begin{equation}}
\newcommand{\ee}{\end{equation}}
\newcommand{\bea}{\begin{eqnarray}}
\newcommand{\eea}{\end{eqnarray}}
\def\nn{\nonumber}

\newcommand{\tvec}[1]{\mbox{\boldmath{$#1$}}}

\def\ket#1{\hbox{$\vert #1\rangle$}}   
\def\bra#1{\hbox{$\langle #1\vert$}}   
\def\oneh{{\textstyle {1\over 2}}}
\def\onet{{\textstyle {1\over 3}}}

\def\oneq{{\textstyle {1\over 4}}}
\def\treh{{\textstyle {3\over 2}}}

\def\oneseven{{\textstyle {1\over 7}}}

\def          
\simeq{{\ \lower2pt\hbox{$-$}\mkern-13mu \raise2pt \hbox{$\sim$}\ }}

\def           
\slasha#1{{\hbox{$\not$} \mkern-3mu\hbox{$\,#1\,$}}}

\begin{document}

\newcommand*{\Pavia}{Dipartimento di Fisica Nucleare e Teorica, 
Universit\`a degli Studi di Pavia, Pavia, Italy}\affiliation{\Pavia}
\newcommand*{\INFN}{Istituto Nazionale di Fisica Nucleare, 
Sezione di Pavia, Pavia, Italy}\affiliation{\INFN}
\title{Parton content of the nucleon from distribution amplitudes and transition distribution amplitudes}

\author{B.~Pasquini}\affiliation{\Pavia}\affiliation{\INFN}
\author{M.~Pincetti}\affiliation{\Pavia}\affiliation{\INFN}
\author{S.~Boffi}\affiliation{\Pavia}\affiliation{\INFN}



\begin{abstract}
The nucleon distribution amplitudes and the nucleon-to-pion transition distribution amplitudes are investigated at leading twist within the frame of a light-cone quark model. The distribution amplitudes probe the three-quark component of the nucleon light-cone wave function, while higher order components in the Fock-space expansion of the nucleon state are essential to describe the nucleon-to-pion transition distribution amplitudes. Adopting a meson-cloud model of the nucleon the nucleon-to-pion transition distribution amplitudes are calculated for the first time.
\end{abstract}
\maketitle


\section{Introduction}
\label{sec:intro}

The hadron structure is believed to be described in terms of the fundamental theory of strong interactions, quantum chromodynamics (QCD), whose equations are notoriously difficult to solve. A successful approach in high-energy scattering is based on light-front quantization where hadrons are described by light-cone wave functions (LCWFs)~\cite{Brodsky:1997de}. The latter are expressed as an expansion of various quark ($q$), antiquark ($\bar q$) and gluon ($g$) Fock components. Schematically, a nucleon state is conceived as the following superposition
\be
\label{eq:fock}
\ket{N} = \psi_{(3q)}\ket{qqq} + \psi_{(3q+1g)}\ket {qqqg} +\psi_{(3q+q\bar q)} \ket{qqqq\bar q} +\dots,
\ee
where in the light-cone gauge $A^+=0$ the valence three-quark LCWF $\psi_{(3q)}$ involves six independent amplitudes corresponding to different combinations of quark orbital angular momentum and helicity~\cite{Ji:2002xn}, and the Fock component $ \psi_{(3q+1g)}$ with three quarks plus one gluon involves 126 independent amplitudes~\cite{Ji:2003yj}. Adding a pair of sea quarks into the valence component to build the amplitude $\psi_{(3q+q\bar q)}$ leads to an even more complicated LCWF.

To probe the parton content of the nucleon suitable models have to be invented to give explicit expressions for the LCWF amplitudes and exclusive processes have to be explored, where a large space-like  momentum is transferred to an intact hadron. The framework for analyzing such processes  was developed more than thirty years ago investigating elastic and inelastic form factors~\cite{Lepage:1979zb,Lepage:1980fj,Efremov: 1979qk} and relies on perturbative QCD (pQCD). According to the factorization theorem, the scattering amplitudes can be expressed as convolutions of the (process-dependent but  perturbatively calculable) hard kernel of the process and the nonperturbative (process independent) contribution describing the hadrons that take part in the reaction. In the case of form factors this contribution is represented by distribution amplitudes (DAs)~\cite{Radyushkin:1977gp,Chernyak:1977as} that describe the hadron structure in parton configurations at small transverse separation. In the nucleon case, DAs probe that part of the nucleon state with orbital angular momentum $L_z=0$ and at leading twist they involve only two of the six amplitudes entering the valence three-quark Fock component $\psi_{(3q)}$.

The properties of DAs were first studied using the method of QCD sum rules developed in Ref.~\cite{Shifman:1978bx}. This method gives the possibility to calculate the values of DA moments in terms of suitable sum rules. Therefore, knowing the behavior of the first few moments one can reconstruct the main properties of DAs as originally shown in Refs.~\cite{Ioffe:1981kw,Chernyak:1983ej,Chernyak:1984bm,Chernyak:1987nt,Chernyak:1987nu}. Although some work is available for other baryons (see, e.g.,~\cite{Chernyak:1987nu,Braun:1999te,Huang:2006ny,Ball:2008fw,Braun:2008ia,Liu:2008yg}), the existing investigations were mainly focused on the nucleon DAs (see~\cite{Chernyak:1983ej} for an early review and~\cite{Braun:2006hn} for a more recent one). Estimates of the nucleon DAs based on QCD sum rules can be found in Refs.~\cite{Gari:1986ue,Gari:1986dr,King:1986wi,Eckardt:1993ur,Eckardt:1993hn,Carlson:1986zs,Stefanis:1987vr,Schaefer:1989cy,Stefanis:1992nw,Bergmann:1993rz,Stefanis:1999wy,Lenz:2009ar}. The nucleon DAs were systematically studied in Ref.~\cite{Braun:2000kw} up to twist six and related to  the nucleon form factors~\cite{Chernyak:1987nv,Braun:2001tj,Braun:2006hz} and the $N\to\Delta$ transition at intermediate values of the momentum transfer~\cite{Lenz:2009ar} using light-cone sum rules. A variety of model calculations~\cite{Dziembowski:1987zq,Schaefer:1989tt,Dziembowski:1990md,Bolz:1996sw} have also been derived from  dynamical or phenomenological Ans\"atze for the nucleon wave function in order to describe the intermediate/low $Q^2$ region where the nonperturbative features of QCD are significant. Valuable additional information is also provided by lattice QCD~\cite{Richards:1986dc,Martinelli:1988xs,Gockeler:2007qs,Braun:2008ur,Gockeler:2008xv}.

In other processes like deeply virtual Compton scattering (DVCS) or hard exclusive meson production the concept of generalized parton distributions (GPDs)~\cite{Muller:1994} has proven to be useful (for reviews, see~\cite{Goeke:2001tz}). GPDs have been introduced as universal nonpertubative objects describing the hadron structure in terms of nondiagonal hadronic matrix elements of bilocal products of the light-front quark and gluon field operators. Their crossed version defines the generalized distribution amplitudes (GDAs) that describe the hadronization of a quark-antiquark or gluon pair in a pair of hadrons, e.g. a pair of $\pi$ mesons, $\gamma^*\gamma\to\pi\pi$~\cite{Diehl:1998dk}. Other matrix elements can be defined generalizing the concept of GPDs for non-diagonal transitions~\cite{Frankfurt:1999fp,Frankfurt:1998jq}  and describing the transition amplitude between two hadrons or between a hadron and a real photon, thus called transition distribution amplitudes (TDAs). 

Recently, attention has been drawn to TDAs under the assumption that the factorization theorems for exclusive processes~\cite{Collins:1996fb} also apply to reaction mechanisms like proton-antiproton annihilation into two photons, $p\bar p\to\gamma^*\gamma$, in the near forward region and large virtual photon invariant mass $Q$~\cite{Pire:2004ie} or into a pion and a high-$Q^2$ lepton pair in the forward region, $p\bar p\rightarrow\gamma^*\pi\rightarrow l^+\,l^-\pi$~\cite{Pire:2005ax}, exclusive meson pair production in $\gamma^*\gamma$ scattering at small momentum transfer~\cite{Lansberg:2006fv}, DVCS on a proton target in the backward region~\cite{Lansberg:2006uh}, or hard exclusive electroproduction of a pion in the backward region, $\gamma^* N\rightarrow N'\pi$~\cite{Lansberg:2007ec}. Within the factorization scheme, the hard and soft subprocesses decouple in the amplitude for these reactions, the soft part being a universal nonperturbative object describing the transition from a hadron to a real photon, or a proton to a pion. 

Depending on the values of the Mandelstam variables $s$ and $t$ in $\gamma^*\gamma$ scattering, a dual factorization mechanism has been identified in Ref.~\cite{Anikin:2008bq} describing the fusion of a real photon  and a highly virtual and longitudinally polarized photon. One mechanism takes place when $s\ll Q^2$, while $t$ is of the order of $Q^2$, and involves twist-three GDAs, whereas the other one occurs for $t\ll Q^2$ and $s\sim Q^2$ and employs leading-twist $\gamma\to\pi$ TDAs. Such TDAs have recently been studied in the Nambu-Jona-Lasinio model~\cite{Courtoy:2008nf}. The $\gamma\to\pi^-$ TDAs are connected to the $\pi^+\to\gamma$ TDAs through CPT symmetry~\cite{Courtoy:2008ij}. The vector and axial-vector $\pi\to\gamma$ TDAs have been analyzed in a quark model~\cite{Tiburzi:2005nj}, in the spectral quark model~\cite{Broniowski:2007fs} and in the Nambu-Jona-Lasinio model~\cite{Courtoy:2007vy}. 

When studying the nucleon structure the $N\to\pi$ TDAs are particularly interesting because they directly probe the three-quark plus sea $q\bar q$ pair component $\psi_{(3q+q\bar q)}$ in Eq.~(\ref{eq:fock}). The possibility to extract experimental information on the $N\to\pi$ TDAs has been studied in Refs.~\cite{Pire:2005ax,Lansberg:2007se} for the $\bar pN\rightarrow \gamma^* \pi$ reaction in the kinematical regime accessible by GSI-FAIR~\cite{Spiller:2006gj} and in Ref.~\cite{Lansberg:2007ec} for the $\gamma^* N\rightarrow N'\pi$ reaction in the kinematical conditions of JLab. In such pioneering works the TDAs were predicted on the basis of the soft-pion theorems~\cite{Polyakov:2006dd} which allow to calculate three out of the eight independent leading-twist TDAs in terms of the proton DAs. Predictions for the TDAs in the soft-pion limit were also obtained in Ref.~\cite{Braun:2007pz}. However, it is desirable to extend these analyses to a more general framework, using as input different model calculations and also going beyond the kinematical soft-pion limit just because the $N\to \pi$ TDAs represent a new tool to access information on the Fock-space components of the nucleon wave function beyond the valence-quark contribution. Furthermore, in the impact parameter representation the $N\rightarrow \pi$ TDAs map out the transverse location of the small-size core and the meson cloud inside the proton~\cite{Pire:2005mt}.
 
Being nonperturbative quantities, DAs, GPDs, GDAs and TDAs cannot be calculated from first principles, but have to be described by models or derived within lattice QCD. In this paper we are concerned with nucleon DAs and $N\to\pi$ TDAs within a phenomenological model for the LCWFs of the nucleon based on the  light-cone constituent quark model (CQM) that has successfully been applied to the calculation of generalized parton distributions~\cite{Boffi:2002yy,Boffi:2003yj,Pasquini:2007xz,Pasquini:2005dk}, and transverse momentum dependent parton distributions~\cite{Pasquini:2008ax,Boffi:2009sh} taking into account the full decomposition of the three-quark Fock component of the nucleon state. In order to derive expressions for the $N\to\pi$ TDAs we have to implement the model in order to include Fock components with a sea $q\bar q$ pair. This will be done assuming the nucleon to consist of a bare three-quark object surrounded by a meson cloud along the lines that were already followed in the calculation of the nucleon GPDs~\cite{Pasquini:2006dv} and electroweak form factors~\cite{Pasquini:2007iz}. This will allow us to give a first estimate of the $N\to\pi$ TDAs for future applications.

The paper is organized  as follows. The nucleon DAs, whose definition and properties are recalled in Sec.~\ref{sec:DA}, are explicitly derived in the light-cone CQM in Sec.~\ref{sec:DALCWF} and numerically computed in Sec.~\ref{sec:DALC}. The $N\to\pi$ TDAs are derived within the meson-cloud model in Sec.~\ref{Sec:TDAs}, and some results are presented in Sec.~\ref{sec:results}. Concluding remarks are given in the final Section. The spin components required by the model for the baryon LCWFs are listed in the Appendix.


\section{Nucleon distribution amplitudes}
\label{sec:DA}

In this Section we recall some important definitions and properties of the nucleon DAs. 

In coordinate space, the proton DAs are derived from the following proton-to-vacuum matrix elements of trilocal operators built of quarks and gluon fields~\cite{Chernyak:1984bm,King:1986wi,Chernyak:1987nu}
\begin{equation}\label{General-Matrix-Element}
\langle 0|\epsilon^{ijk}u^{i'}_\alpha(z_1n)[z_1;z_0]_{i'i}
u^{j'}_\beta(z_2n)[z_2;z_0]_{j'j}
d^{k'}_\gamma(z_3n)[z_3;z_0]_{k'k}|p(p_p,\lambda)\rangle,
\end{equation}
where $\ket{p(p_p,\lambda)}$ denotes the proton state with momentum $p_p$ ($p_p^2=M^2$) and helicity $\lambda$; $u$, $d$ are the field operators for up and down quarks, respectively; the Greek letters $\alpha$, $\beta$ and $\gamma$ stand for Dirac indices, while the Latin letters $i$, $j$ and $k$ refer to color; $n$ is an arbitrary light-like vector ($n^2=0$) and $z_i$ are real numbers that specify quark separation, with $\sum_iz_i=1$. In Eq.~(\ref{General-Matrix-Element}) the gauge factors $[z_i ; z_0]$ render the matrix element gauge-invariant and are defined as
\be
\label{wilson}
[z_i ; z_0] \equiv \mathcal{P}\exp\Big[ig(z_i - z_0)\int_0^1dt\,n_\mu A^\mu\Big(n[tz_i + (1 - t)z_0]\Big)\Big],
\ee
where $\mathcal{P}$ indicates the path-ordering prescription. In the following we will work in the light-cone gauge  $A^+ = 0$ where the gauge factors reduce to the identity. 

Taking into account Lorentz covariance, spin, and parity conservation of the nucleon, the most general decomposition of the matrix element in Eq.~(\ref{General-Matrix-Element}) involves 24 invariant 
functions~\cite{Braun:2000kw}. To the leading twist-three accuracy only three of them are relevant, 
and are given by the Lorentz invariant (scalar) functions of positive parity $V$(= vector), $A$(= axial-vector), and $T$(= tensor):
\begin{eqnarray}
\label{DA}
\bra{0}\epsilon^{ijk}u^i_\alpha(z_1n) u^j_\beta(z_2n)d^k_\gamma(z_3n)\ket{p(p_p,\lambda)} 
&=&
\oneq f_N\Big[(\slashed{p}C)_{\alpha\beta}(\gamma_5N^+)_\gamma V(z_i n\cdot p) 
+ (\slashed{p}\gamma_5C)_{\alpha\beta}(N^+)_\gamma A(z_i n\cdot p) \nonumber\\
&&{} +
(\sigma_{p\mu}C)_{\alpha\beta}(\gamma^\mu\gamma_5N^+)_\gamma T(z_i n\cdot p)\Big],
\end{eqnarray}
where $\sigma^{\mu\nu} = \oneh[\gamma^\mu, \gamma^\nu]$,  $\sigma^{p\mu}$ is a shorthand notation for $p_\nu\sigma^{\nu\mu}$, $C$ is the charge conjugation matrix and $N^+$ is the light-cone \lq\lq good'' or \lq\lq large'' component of the nucleon spinor $N$. In Eq.~(\ref{DA}) the proton decay constant $f_N$ is a dimensional quantity representing the value of the nucleon distribution amplitude at the origin of the configuration space~\cite{Chernyak:1983ej,Chernyak:1984bm,Ioffe:1981kw}. Furthermore, we introduced a second light-like vector $p^\mu$ such that $2 p\cdot n=1$. In particular, we make the following choice for the vectors $p^\mu$ and $n^\mu$:
\be
p^\mu=\frac{p_p^+}{\sqrt{2}}(1,0,0,1),
\qquad n^\mu=\frac{1}{2\sqrt{2}p_p^+}(1,0,0,-1).
\ee

It is convenient to define the functions $V$, $A$ and $T$ in momentum space
\be
\label{eq:fourier}
V(x_1,x_2,x_3) = (n\cdot p)^3\int\prod_{j=1}^3\frac{\mathrm{d}z_j}{(2\pi)^3} 
V(z_1 n\cdot p,z_2 n\cdot p,z_3 n\cdot p)
\exp\Big[i\sum_{k=1}^3 x_kz_k(n\cdot p)\Big],
\ee
and similarly for $A$ and $T$. The variables $x_i$ conjugate to the light-cone positions of the quark operators in (\ref{DA}) are collinear momentum fractions  of the proton longitudinal momentum carried by each quark  in the infinite momentum frame, with $0\leq x_i\leq 1$ and $\sum_{i=1}^3x_i=1$ by momentum conservation. Accordingly, the scalar  functions $V(x_i)$, $A(x_i)$ and $T(x_i)$ are DAs describing the longitudinal momentum distributions in the nucleon at a fixed scale $\mu^2$ that is not explicitly indicated if not necessary. 

Because of permutation symmetry between the two up quarks, the functions $V$ and $T$ are symmetric and $A$ antisymmetric in their first two arguments. In addition, the requirement that the three quarks have to be coupled to give an isospin $\oneh$ state (the nucleon), yields the relations
\bea
\label{rel1}
2 T(1,2,3)&=& \Phi(1,3,2)+\Phi(2,3,1),
\\
\label{rel2}
\Phi(1,2,3)&=& V(1,2,3)-A(1,2,3),
\eea
which allow to express the proton DAs in terms of a single independent scalar function $\Phi$ with mixed symmetry.

Introducing quark fields with definite chirality and denoting the Fourier transform of the matrix element on the left-hand side of Eq.~(\ref{DA}) by $D^\lambda_{\alpha\beta,\gamma}$, the three DAs can be obtained as
\bea
\label{V1}
V &=& \frac{1}{(2)^{1/4}(p_p^+)^{3/2}\,f_N }\Big(D_{12,1}^\uparrow +D_{21,1}^\uparrow\Big),
 \\
\label{A1}
A &=& -\frac{1}{2^{1/4}(p_p^+)^{3/2}\,f_N } \Big(D_{12,1}^\uparrow - D_{21,1}^\uparrow\Big),
 \\
\label{T1}
T &=& -\frac{1}{2^{1/4}(p_p^+)^{3/2}\,f_N } \,D_{11,2}^\uparrow,
\eea
where the $\uparrow,\downarrow$ arrows denote the up and down helicity of the proton, respectively.
Thus,
\be
\label{Phi}
\Phi = \frac{2}{(2)^{1/4}(p_p^+)^{3/2}\,f_N } D_{12,1}^\uparrow.
\ee

Eq.~(\ref{DA}) is equivalent to writing the three-quark $uud$ component of the proton state with positive helicity in the infinite momentum frame as~\cite{Chernyak:1984bm,Chernyak:1987nt,Chernyak:1987nu}~\footnote{%
Note that here the $uud$ component is normalized as
$$
{}_{uud}\bra{p,\lambda} p',\lambda'\rangle_{uud} 
= \onet 2(2\pi)^3p^+\delta(p'^+-p^+)\delta^{(2)}({\mathbf p}'_\perp-{\mathbf p}_\perp)\delta_{\lambda'\lambda} .
$$
}%
\bea
\ket{p(p_p,\uparrow)}_{uud}
&=& \frac{1}{\sqrt{3}} \frac{f_N}{4}\int_0^1\left[\frac{{\rm d}x}{\sqrt{x}}\right]_3
\left\{\frac{V-A}{2}\ket{u^\uparrow(\tilde k_1)u^\downarrow(\tilde k_2)d^\uparrow(\tilde k_3)}\right.\nn\\
&&{}+ \left.
\frac{V+A}{2}\ket{u^\downarrow(\tilde k_1)u^\uparrow(\tilde k_2)d^\uparrow(\tilde k_3)}
-T \ket{u^\uparrow(\tilde k_1)u^\uparrow(\tilde k_2)d^\downarrow(\tilde k_3)}\right\},
\label{DA-proton-state}
\eea
or in a more compact way as
\bea
\ket{p(p_p,\uparrow)}_{uud}&=&\frac{1}{\sqrt{3}} \frac{f_N}{4}\int_0^1\left[\frac{{\rm d}x}{\sqrt{x}}\right]_3
\Phi
\left[\ket{u^\uparrow(\tilde k_1)u^\downarrow(\tilde k_2)d^\uparrow(\tilde k_3)}-\ket{u^\uparrow(\tilde k_1)d^\downarrow(\tilde k_2)u^\uparrow(\tilde k_3)} \right],
\label{DA-proton-state2}
\eea
where the integration measure is defined as
\be
\label{eq:19}
\left[\frac{{\rm d} x}{\sqrt{x}}\right]_N 
= \left(\prod_{i=1}^N \frac{{\rm d} x_i}{\sqrt{x_i}}\right) \delta\left(1-\sum_{i=1}^N x_i\right).
\ee
In Eqs.~(\ref{DA-proton-state}) and (\ref{DA-proton-state2}) 
\be
\ket{u^\uparrow(\tilde k_1)u^\downarrow(\tilde k_2)d^\uparrow(\tilde k_3)}
= \frac{\epsilon^{ijk}}{\sqrt{6}} b^{\dagger i}_u(\tilde k_1,\uparrow) b^{\dagger j}_u(\tilde k_2,\downarrow) b^{\dagger k}_d(\tilde k_3,\uparrow) \ket{0},
\ee
where $b^{\dagger c}_{u,d}(\tilde k,\lambda)$ are the creation operators of free $u$ and $d$ quarks with momentum $\tilde k=(k^+,\mathbf{k}_{\perp})$ ($k^+=(k^0+k^3)/\sqrt{2}$ and $\mathbf{k}_{\perp}$ being the plus and transverse momentum components, respectively), helicity $\lambda$ and color $c$ (see also Eq.~(\ref{eq:threequark}) below).

The corresponding neutron state is obtained from (\ref{DA-proton-state2}) by interchanging $u$ and $d$, with an overall change of sign.

The matrix elements $D^\lambda_{\alpha\beta,\gamma}$, and ultimately the DAs, are directly linked to the $L_z=0$ component of the valence-quark wave function of the nucleon, by integrating out the transverse momenta of the constituent quarks~\cite{Chernyak:1987nt,Chernyak:1987nu,Dziembowski:1987zq,Dziembowski:1990md} (see Eqs.~(\ref{final}) and (\ref{Phi-final}) below).

The DAs for the nucleon are well known in two limits~\cite{Chernyak:1984bm}. The first is  the static SU(6) symmetric quark model, where the variables $x_i$ take on only discrete values and the distribution amplitude is totally symmetric:
\be
\label{Phi_nr}
\Phi_{{\rm NR}}=\delta(x_1-\onet)\delta(x_2-\onet)\delta(x_3-\onet).
\ee
The second is the asymptotic regime of sufficiently high $Q^2$ where $\Phi$ takes the form
\be
\label{Phi_as}
\Phi_{{\rm AS}}=120x_1x_2x_3,
\ee
which is totally symmetric under quark exchange and has the flavor-spin structure assumed in the SU(6)-symmetric quark model. In the asymptotic limit ($Q^2\to\infty$), $A$ becomes negligible because of the Pauli principle, and $V$ and $T$ become totally symmetric under particle exchange, i.e. $V,T\to \Phi_{{\rm AS}}$~\cite{Stefanis:1987vr}.

Both limits (\ref{Phi_nr}) and (\ref{Phi_as}) are conflicting with experiment. In the first case one obtains wrong results for the neutron and proton magnetic form factors ($G_M^n>0$, $G_M^p<0$), in the second case $G_M^n>0$ and $G_M^p/G_M^n\to 0$ at large $Q^2$~\cite{Chernyak:1984bm}.

At intermediate values of $Q^2$ the nucleon DAs turn out to be quite different from their nonrelativistic and asymptotic limits. The analysis takes advantage of moments of the DA $\Phi$ defined as
\be
\phi^{(l,m,n)}= \frac{1}{\phi_N}
\int [{\rm d}x]\, x_1^l x_2^m x_3^n\, \Phi(x_1,x_2,x_3),
\ee
where $[{\rm d}x]={\rm d}x_1{\rm d}x_2{\rm d}x_3\delta(1-x_1-x_2-x_3)$, and $\phi_N$ is a normalization constant, which is chosen such that $\phi^{(0,0,0)}=1$.
Longitudinal momentum conservation ($x_1+x_2+x_3=1$) imposes the following condition 
\begin{eqnarray}
\phi^{(n_1,n_2,n_3)}=\phi^{(n_1+1,n_2,n_3)}+\phi^{(n_1,n_2+1,n_3)}
+\phi^{(n_1,n_2,n_3+1)}.
\end{eqnarray}
Thus, not all the moments at a given order $M=n_1+n_2+n_3$ 
are linearly independent. 

The DA  moments can be expressed in terms of matrix elements of suitable local operators entering appropriate sum rules~\cite{Chernyak:1981zz,Chernyak:1984bm,Chernyak:1987nu} following the lines of the method of QCD sum rules developed in Ref.~\cite{Shifman:1978bx}.

The nucleon DA obeys a renormalization-group equation which requires that $\Phi(x_i,Q^2)$ is only logarithmically dependent on the momentum transfer scale $Q^2$~\cite{Lepage:1980fj}.
Following Refs.~\cite{Stefanis:1999wy,Stefanis:1994zd}, the scale dependence of the nucleon DA can be cast in the form
\be
\label{evolution}
\Phi(x_i,Q^2)=\Phi_{{\rm AS}}(x_i)
\sum_{n=0}^\infty B_n(\mu^2) \tilde\Phi_n(x_i)
\left[\frac{\log(Q^2/\Lambda_{{\rm QCD}}^2)}{\log(\mu^2/\Lambda_{{\rm QCD}}^2)}\right]^{-\gamma_n},
\ee
where $\tilde\Phi_n(x_i)$ are orthogonal eigenfunctions of the nucleon evolution equations, orthonormalized within a basis of Appell polynomials, and $\gamma_n$ are the anomalous dimensions listed in Ref.~\cite{Lepage:1980fj}. In Eq.~(\ref{evolution}) the projection coefficients $B_n(\mu^2)$ encode the non-perturbative input of the bound state dynamics at the factorization (renormalization) scale $\mu^2$. Using  the explicit expression for the eigenfunctions  $\tilde\Phi_n$ in terms of Appel polynomials they can be expressed as linear combinations of DA moments, i.e. 
\be
\label{appel-coeff}
B_n(\mu^2)=\frac{N_n}{120}\sum_{i,j=0}^n a_{ij}^n\phi^{(i,0,j)}(\mu^2),
\ee
where the coefficients $a_{ij}^n$ and the normalization constant $N_n$ have been calculated up to order $M=i+j=9$ in Refs.~\cite{Bergmann:1994mz,Stefanis:1999wy}.

An alternative expansion of DAs is possible in terms of contributions of operators with a given conformal spin~\cite{Braun:1999te,Braun:2008ia,Braun:2000kw,Braun:2006hz}. This is convenient since operators with different spin do not mix under renormalization in one loop, and only operators with the same spin can be related by equations of motion so that the truncation of the conformal spin expansion at a certain order produces a self-consistent approximation. For example, at leading twist with the minimum possible conformal spin, $\Phi$ reduces to $\Phi_{\rm AS}$, whereas its conformal expansion to the next-to-leading conformal spin accuracy reads~\cite{Braun:2000kw,Braun:2006hz}
\be
\Phi(x_i,\mu^2) = \Phi_{\rm AS}(x_i)\phi_3^0(\mu^2)\left[1+\tilde\phi_3^-(\mu^2)(x_1-x_2) + \tilde\phi_3^+(\mu^2)(1-3x_3)\right],
\ee
where 
\be
\phi_3^0 = f_N , \quad \tilde\phi_3^- =\frac{21}{2}\left[\phi^{(1,0,0)} - \phi^{(0,1,0)}\right], \quad
\tilde\phi_3^+ = \frac{7}{2}\,\phi^{(0,0,1)}.
\ee
Numerical estimates of the coefficients at $\mu^2=1$ GeV$^2$ available from QCD sum rules~\cite{Chernyak:1984bm,Chernyak:1987nt,King:1986wi,Chernyak:1987nv} give
\be
f_N = (5.3\pm 0.5)\times 10^{-3}\ {\rm GeV}^2,\quad 
\tilde\phi_3^- = 4.0\pm 1.5, \quad
\tilde\phi_3^+ = 1.1\pm 0.3.
\label{eq:fn}
\ee
An approximately 40\% lower value of $f_N$ has been determined recently in lattice calculations~\cite{Braun:2008ur}: $f_N= 3.234(63)(86)\times 10^{-3}$ GeV$^2$ at $\mu^2=1$ GeV$^2$.
In our numerical estimates the value in Eq.~(\ref{eq:fn}) will be used.

The resulting DA exhibits a broad and rich structure that is reflected in an asymmetric distribution of the proton momentum between the three valence quarks (in the limit of infinite momentum). According to the QCD sum rule approach about 60\% of the proton longitudinal momentum is carried by the up quark with its helicity in the same direction as that of the proton. The remaining up and down quarks, with combined helicity zero, are confined into the  small-$x$ region, each carrying about 20\% of the total longitudinal momentum. This asymmetry is a common feature of all octet baryons~\cite{Chernyak:1987nu}. A somewhat smaller asymmetry of the helicity amplitude $\uparrow\downarrow\uparrow$ for the $uud$ configuration was found in Ref.~\cite{King:1986wi} re-evaluating the momentum sum rules. This suggests the possibility of spin-zero diquark clustering in the nucleon wave function as a manifestation of the attractive QCD force produced by gluon exchange that is just strongest in the spin-zero quark-quark state~\cite{Dziembowski:1990md}. However, lattice QCD calculations of the first two moments were unable to confirm this asymmetric behavior of the nucleon DAs~\cite{Martinelli:1988xs}.

The problem with QCD sum rules is that the moment sum rules are not stringent enough to fix the shape of the nucleon DAs uniquely~\cite{Gari:1986dr,Schaefer:1989cy}. With increasing order of expansion the oscillations become stronger, and small variations of the moments may lead to a completely artificial behavior~\cite{Eckardt:1993hn}. Such an extreme sensitivity of the expansion coefficients indicates that the moments do not give a convergent expansion. Actually, there is an infinite number of possible solutions which satisfy the moment sum rules, but differ dramatically in their shape. Thus, the predicted observables, like form factors, may result quite different. To determine the possible variation of DAs allowed by moment sum rules and to reconcile the constraints from moment sum rules with data, the heterotic solution was proposed in Refs.~\cite{Stefanis:1987vr,Stefanis:1992nw} combining $Q^2$ evolution equations in pQCD, QCD sum rules and phenomenology. Allowing some flexibility to the expansion coefficients (\ref{appel-coeff}), while keeping them as close as possible to the sum-rule requirements, a good agreement with high-$Q^2$ data on the proton magnetic form factor was achieved.

Alternative phenomenological approaches take advantage of constraints on 
the three-quark component of the nucleon wave function imposed by data. 
For example, requiring that with the same nucleon wave function one 
reproduces the proton form factors, the phenomenological valence quark 
distribution as well as the $J\psi\to p\bar p$ decay width, the authors 
of Ref.~\cite{Bolz:1996sw} assumed a wave function with a small hard 
factorizing part depending on the longitudinal momenta and described by 
a nucleon DA and a large soft nonfactorizing contribution depending on 
$x_i$ and $\mathbf k_{i\perp}$ solely in the combination 
$k^2_{i\perp}/x_i$ with a Gaussian fall-off with $k_{i\perp}$. All 
requirements are met with a model wave function depending on only two 
parameters, namely the proton decay constant $f_N$ and the size 
parameter of the transverse momentum dependence. This gives a DA that is 
much less asymmetric than that derived from QCD sum rules, rather 
resembling the asymptotic DA, but with the position of the only maximum 
somewhat shifted.
 Actually, the asymmetry of the leading-twist amplitude constraint by phenomenology~\cite{Bolz:1996sw,Braun:2000kw,Braun:2006hz} or calculated on the lattice~\cite{Gockeler:2007qs,Braun:2008ur,Gockeler:2008xv} is much smaller than in QCD sum rule calculations (see Table~\ref{Moments_comparison} discussed in Sec.~\ref{sec:DALC}).


\section{Nucleon distribution amplitudes and light-cone wave functions}
\label{sec:DALCWF}

In this Section we derive explicit expression for the matrix elements $D^\lambda_{\alpha\beta,\gamma}$ in  terms of LCWFs. To this aim, we first introduce the Fourier expansion in momentum space of
the quark field operator of flavour $q$ and colour $c$~\cite{Lepage:1980fj,Brodsky:1997de}
\begin{eqnarray}
\label{Free-quark-field}
q^c(zn^-, z\mathbf{n}_\perp) 
&=&
\int\frac{dk^{+}d^2\mathbf{k}_\perp}{16\pi^3k^{+}}\Theta(k^+)
\nonumber\\
&& {}\times\sum_\lambda\{
b^c_q(\tilde k,\lambda)u_+(\tilde k,\lambda)\exp(-ik^+zn^- + i\mathbf{k}_\perp \cdot z\mathbf{n}_\perp)
\nonumber\\
&&{}  + d^{\dagger c}_q(\tilde k, \lambda)v_+(\tilde k,\lambda)\exp(+ik^+zn^- - i\mathbf{k}_\perp \cdot
z\mathbf{n}_\perp)\},
\end{eqnarray}
where the $b$ and $d^\dagger$ operators respectively annihilate the  \lq\lq good'' component of the quarks fields and create the \lq\lq good'' component of the
antifields fulfilling the anticommutations relations
\begin{eqnarray}\label{Anticommutations-rules}
\{b^{c'}_{q'}(\tilde k', \lambda'), b^{\dagger
c}_q(\tilde k, \lambda)\} & = &\{d^{c'}_{q'}(\tilde k', \lambda'), d^{\dagger
c}_q(\tilde k, \lambda)\}\nonumber\\
& = & 16\pi^{3}k^+\delta(k'^+ - k^+)\delta^{(2)}(\mathbf{k'}_\perp -
\mathbf{k}_\perp)\delta_{q'q}\delta_{\lambda'\lambda}\delta_{c'c}.
\end{eqnarray}
In Eq.~(\ref{Free-quark-field}), $u_+(\tilde k,\lambda)$ and $v_+(\tilde k,\lambda)$
are the light-cone spinors of the quark and antiquark, respectively.

The three-quark Fock component of the light-front proton state is given by
\be
\ket{p(p_p,\lambda)} = \sum_{\tau_i,\lambda_i,c_i}
\int\left[\frac{{\rm d}\xi }{\sqrt{\xi}}\right]_3
[{\rm d}^2{\bf k}_\perp]_3
\Psi_\lambda^{p,[f]}(\{\xi_i,{\bf k}_{i\perp };\lambda_i,\tau_i\}_{i=1,2,3})
\prod_{i=1}^3
\ket{q^{\lambda_i};\xi_i p^+_p, \, {\bf p}_{i\perp }},
\label{eq:18}
\ee
where $\Psi_\lambda^{p,[f]}(\{\xi_i,{\bf k}_{i \perp };\lambda_i,\tau_i\}_{i=1,2,3})$ is the momentum LCWF which gives the probability amplitude for finding in the nucleon three quarks with momenta
 $(\xi_i p^+_p, {\bf p}_{i\perp}= {\bf k}_{i\perp}+\xi_i p^+_p)$, and spin and isospin variables $\lambda_i$ and $\tau_i,$ respectively. The proton state is normalized as
\be
\langle p,\lambda|p',\lambda\rangle=2(2\pi)^3p^+\delta(p'^+-p^+)\delta^{(2)}({\bf p}'_\perp-{\bf p}_\perp),
\ee
and the three-quark state is defined as
\be
\label{eq:threequark}
\prod_{i=1}^3
\ket{q^{\lambda_i};\tilde k_i}=
\frac{\epsilon^{ijk}}{\sqrt{6}}b_q^{\dagger i}(\tilde k_1,\lambda_1)
b_q^{\dagger j}(\tilde k_2,\lambda_2)
b_q^{\dagger k}(\tilde k_3,\lambda_3)|0\rangle.
\ee
In Eq.~(\ref{eq:18}) and in the following formulas,  the integration measures are defined by (\ref{eq:19}) and
\be
\label{eq:20}
[{\rm d}^2{\bf k}_\perp]_N = \left(\prod_{i=1}^N
\frac{{\rm d}^2{\bf k}_{i\perp}}{2(2\pi)^3}\right)\,2(2\pi)^3\,
\delta\left(\sum_{i=1}^N {\bf k}_{i\perp}\right).
\ee

Inserting the momentum-space expansion (\ref{Free-quark-field}) of the quark fields and the proton Fock-state (\ref{eq:18}) in Eq.~(\ref{DA}),
and using the anticommutation relations for the quark creation 
and annihilation operators, one obtains for the matrix elements 
$D^\lambda_{\alpha\beta,\gamma}$ 
\begin{eqnarray}
&&D^{\lambda}_{\alpha\beta,\gamma}=-24\frac{1}
{\sqrt{x_1x_2x_3}}
u_{+\alpha}(x_1p_p^+,\lambda_1)
u_{+\beta}(x_2p_p^+, \lambda_2)
u_{+\gamma}(x_3p_p^+,\lambda_3)
\nonumber\\
& &\times
\int
[\mathrm{d}^2\mathbf{k}_{\perp}]_3
\Psi_{\lambda}^{p,[f]}\bigg(\{
x_1, \mathbf{k}_{1\perp}; \lambda_{1}, 1/2\},
\{x_2,\mathbf{k}_{2\perp}; \lambda_{2}, 1/2\},
\{x_{3},\mathbf{k}_{3\perp}; \lambda_{3},-1/2\}\bigg).
\label{final}
\end{eqnarray}

The light-cone spinors in Eq.~(\ref{final}) are explicitly given by
\begin{equation} 
u_+(x_ip_p^+,
\uparrow)=\sqrt{\frac{x_ip_p^+}{\sqrt{2}}}\left(\begin{array}{c}
1\\
0\\
1\\
0
 \end{array}
\right)
    \;\;\;\;\;\mathrm{and}\;\;\;\;\;
u_+(x_ip_p^+,
\downarrow)=\sqrt{\frac{x_ip_p^+}{\sqrt{2}}}\left(\begin{array}{c}
0\\
1\\
0\\
-1
\end{array}
\right),\;\;\;i= 1, 2, 3.
\label{LC-spinors}
\end{equation}
As a consequence, the Dirac component of the quark 
spinor selects the quark-spin configuration, with $\alpha=1,3$ 
($\alpha=2,4$) 
corresponding to helicity $\lambda=\uparrow$ ($\lambda=\downarrow$)
of the quark. In particular,
from Eqs.~(\ref{V1})-(\ref{T1})  we see that for a proton 
with helicity $\uparrow$ the DAs are obtained from the wave-function component 
with total quark helicity $\oneh$, corresponding to the projection onto the partial wave with orbital angular momentum  $L_z=0$, and to the three spin configurations
$(\uparrow \uparrow \downarrow)$, 
$(\uparrow \downarrow \uparrow)$, and 
$(\downarrow \uparrow \uparrow)$.
Furthermore, the quantum numbers for the quark isospin in the LCWF in~(\ref{final}) correspond to the isospin projection in the $uud$ configuration.


\section{Nucleon distribution amplitudes in a light-cone quark model}
\label{sec:DALC}

In this Section we estimate the DAs using a phenomenological model~\cite{Boffi:2002yy,Boffi:2003yj,Pasquini:2007xz,Pasquini:2005dk,Pasquini:2008ax,Boffi:2009sh} with LCWFs built in such a way as to satisfy Poincar\'e covariance and to be eigenstates of the total angular momentum operator in the light-front dynamics.
These properties can be fulfilled by constructing the wave function as the product of a momentum-dependent wave function $\tilde \psi(\{x_i,{\bf k}_{i\perp}\})$ which is spherically symmetric and invariant under permutations, and a spin and isospin wave function which is uniquely determined by symmetry requirements and invariant under permutations, i.e.
\begin{eqnarray}
\Psi^{p,[f]}_\lambda(\{x_i, {\bf k}_{i\perp}; \lambda_i, \tau_i\}) 
&=& \tilde \psi(\{x_i,{\bf k}_{i\perp}\})
\nonumber\\
&&{}\times\sum_{\mu_1\mu_2\mu_3}D_{\mu_1\lambda_1}^{1/2*}(R_{cf}(\tilde k_1))D_{\mu_2\lambda_2}^{1/2*}(R_{cf}(\tilde k_2))D_{\mu_3\lambda_3}^{1/2*}(R_{cf}(\tilde k_3))\nonumber\\
   &&{}\times
   \Phi_{\lambda\oneh}^{p}(\mu_1,\mu_2,\mu_3,\tau_1,\tau_2,\tau_3),
\label{InstantFormLINK}
\end{eqnarray}
where $D_{\lambda\mu}^{1/2}(R_{cf}(\tilde{k}))$ 
are matrix elements of the Melosh rotation
$R_{cf}$~\cite{Melosh:1974cu}, which converts the rest-frame spin of the quarks
into light-cone spins.
They are explicitly given by
\begin{eqnarray}
D_{\lambda\mu}^{1/2}(R_{cf}(\tilde k) &=&
\langle\lambda|R_{cf}(\tilde k)|\mu\rangle\nonumber\\
 &=&
\langle\lambda|\frac{m + xM_0 -
i\mathbf{\sigma}\cdot(\hat{\mathbf{z}}\times\mathbf{k}_\perp)}{\sqrt{(m
+ xM_0)^2 + \mathbf{k}_\perp^2}}|\mu\rangle,
\end{eqnarray}
where $m$ is the quark mass and $M_0$ is the mass of the non-interacting three quark system.
In Eq.~(\ref{InstantFormLINK}) the spin and isospin quantum number of the quarks are coupled by the SU(6) symmetric function  $\Phi^p_{\lambda\tau}$
defined as
\begin{equation}
\Phi_{\lambda\tau}^{p}(\mu_1,\mu_2,\mu_3,\tau_1,\tau_2,\tau_3) =
\frac{1}{\sqrt{2}}\Big[\tilde{\Phi}^{0}_{\oneh\lambda}(\mu_1, \mu_2,
\mu_3)\tilde{\Phi}^{0}_{\oneh\tau}(\tau_1, \tau_2, \tau_3) +
\tilde{\Phi}^{1}_{\oneh\lambda}(\mu_1, \mu_2,
\mu_3)\tilde{\Phi}^{1}_{\oneh\tau}(\tau_1, \tau_2, \tau_3)\Big],
\end{equation}
where
\begin{eqnarray}
\label{PHI tilde}
\tilde{\Phi}^{J_{12}}_{J\lambda} = \sum_{M_{J_{12}}}\langle1/2,
\mu_1; 1/2, \mu_2|J_{12}, M_{J_{12}}\rangle\langle J_{12}, M_{J_{12}};
1/2, \mu_3| J, \lambda\rangle.
\end{eqnarray}
In the case of the proton DAs, we need the $uud$ isospin projection of the nucleon 
wave function, which corresponds to 
\begin{eqnarray}
\label{LCWFuud}
\Psi^{p,[f]}_\lambda(\{x_i, {\bf k}_{i\perp}; \lambda_i\}, \{uud\}) 
&=& \frac{1}{\sqrt{2}}\tilde \psi(\{x_i,{\bf k}_{i\perp}\})
\Xi_{\lambda}^{p}(\lambda_1,\lambda_2,\lambda_3)
\tilde{\Phi}^{1}_{\oneh\oneh}(\oneh,\oneh,-\oneh),
\eea
with the isospin coefficient $\tilde{\Phi}^{1}_{\oneh\oneh}(\frac{1}{2},\frac{1}{2},-\frac{1}{2} )
=\sqrt{\frac{2}{3}}$ and the spin dependent part given by
\begin{eqnarray}
\Xi_{\lambda}^{p}(\lambda_1,\lambda_2,\lambda_3)=\sum_{\mu_1\mu_2\mu_3}
D_{\mu_1\lambda_1}^{1/2*}(R_{cf}(\tilde k_1))
D_{\mu_2\lambda_2}^{1/2*}(R_{cf}(\tilde k_2))
D_{\mu_3\lambda_3}^{1/2*}(R_{cf}(\tilde k_3))
\tilde{\Phi}^{1}_{\oneh\lambda}(\mu_1, \mu_2,\mu_3).
\end{eqnarray}

\begin{table}[ht]
\caption{Results for the moments $\phi^{(l,m,n)}$ with $l + m + n \leq 3$ of the
proton DA in different model calculations:
COZ from Ref.~\cite{Chernyak:1987nv}; KS from 
Ref.~\cite{King:1986wi}, SB from Ref.~\cite{Stefanis:1992nw},
DF from Ref.~\cite{Dziembowski:1990md}, BK from Ref.~\cite{Bolz:1996sw},  
and PPB from the present model.}
\label{Moments_comparison}
\begin{center}\footnotesize
\begin{tabular}{ccccccc}
\hline\hline\\
($l$,\,$m$,\,$n$) & $\, $ COZ$\, $ &$\, $ KS$\, $ & $\, $SB$\, $ & $\, $DF$\, $ & $\, $BK$\, $  & $\, $PPB$\, $ \\ \\
\hline\\
$(0\;\; 0\;\; 0)\,$ & $\,    1    \,$  & $\,    1    \,$ & $\, 1   \,$ & $\, 1   \,$ & $\,  1  \,$  & $\,  1  \,$ \\
$(1\;\; 0\;\; 0)\,$ & $\,0.54-0.62\,$  & $\,0.46-0.59\,$ & $\,0.572\,$ & $\, 0.582 \,$ & $\,0.381\,$ & $\,0.346\,$  \\
$(0\;\; 1\;\; 0)\,$ & $\,0.18-0.20\,$  & $\,0.18-0.21\,$ & $\,0.184\,$ & $\, 0.213 \,$ & $\,0.309\,$ & $\,0.331\,$  \\
$(0\;\; 0\;\; 1)\,$ & $\,0.20-0.25\,$  & $\,0.22-0.26\,$ & $\,0.244\,$ & $\, 0.207 \,$ & $\,0.309\,$ & $\,0.323\,$ \\
$(2\;\; 0\;\; 0)\,$ & $\,0.32-0.42\,$  & $\,0.27-0.37\,$ & $\,0.338\,$ & $\, 0.367 \,$ & $\,0.179\,$ &  $\,0.151\,$\\
$(0\;\; 2\;\; 0)\,$ & $\,0.065-0.088\,$  & $\,0.08-0.09\,$ & $\,0.066\,$ & $\, 0.085 \,$ & $\,0.125\,$ & $\,0.141\,$ \\
$(0\;\; 0\;\; 2)\,$ & $\,0.09-0.12\,$  & $\,0.10-0.12\,$ & $\,0.170\,$ & $\, 0.083 \,$ & $\,0.125\,$ & $\,0.136\,$ \\
$(1\;\; 1\;\; 0)\,$ & $\,0.08-0.10\,$  & $\,0.08-0.10\,$ & $\,0.139\,$ & $\, 0.108 \,$ & $\,0.101\,$ &  $\,0.099\,$ \\
$(1\;\; 0\;\; 1)\,$ & $\,0.09-0.11\,$  & $\,0.09-0.11\,$ & $\,0.096\,$ & $\, 0.106 \,$  & $\,0.101\,$& $\,0.096\,$ \\
$(0\;\; 1\;\; 1)\,$ & $\,-0.03-0.03\,$  & $\,\textrm{unreliable}\,$ & $\, -0.021 \,$ & $\,0.018\,$ & $\,0.083\,$ & $\,0.091\,$ \\
$(3\;\; 0\;\; 0)\,$ & $\,0.21-0.25\,$  & $\,\,$ & $\,0.21\,$& $\, 0.249 \,$ & $\,0.095\,$  & $\,0.076\,$\\
$(0\;\; 3\;\; 0)\,$ & $\,0.028-0.04\,$  & $\,\,$ & $\,0.039\,$ & $\, 0.041 \,$ & $\,0.059\,$ & $\,0.070\,$\\
$(0\;\; 0\;\; 3)\,$ & $\,0.048-0.056\,$  & $\,\,$ & $\,0.139\,$ & $\, 0.040 \,$ & $\,0.059\,$ & $\,0.067\,$\\
$(2\;\; 1\;\; 0)\,$ & $\,0.041-0.049\,$  & $\,\,$ & $\,0.079\,$ & $\, 0.060 \,$ & $\,0.042\,$ & $\,0.038\,$\\
$(2\;\; 0\;\; 1)\,$ & $\,0.044-0.055\,$  & $\,\,$ & $\,0.049\,$ & $\, 0.059 \,$ & $\,0.042\,$ & $\,0.037\,$\\
$(1\;\; 2\;\; 0)\,$ & $\,0.027-0.037\,$  & $\,\,$ & $\,0.050\,$ & $\, 0.040 \,$ & $\,0.036\,$ & $\,0.037\,$\\
$(1\;\; 0\; \;2)\,$ & $\,0.037-0.0434\,$  & $\,\,$ & $\,0.037\,$ & $\, 0.039 \,$ & $\,0.036\,$  & $\,0.035\,$\\
$(0\;\; 2\;\; 1)\,$ & $\,-0.004-0.007\,$  & $\,\,$ & $\,-0.023\,$ & $\, 0.004 \,$ & $\,0.030\,$ & $\,0.034\,$\\
$(0\;\; 1\;\; 2)\,$ & $\,-0.005-0.008\,$  & $\,\,$ & $\,-0.007\,$ & $\, 0.005 \,$ & $\,0.030\,$ & $\,0.033\,$\\
\\
\hline\hline
\end{tabular}
\end{center}
\end{table}

Inserting the wave function~(\ref{LCWFuud}) in Eq.~(\ref{final}), we find for the nucleon DAs the 
final results
\begin{eqnarray}
\label{V-final}
V&=&
-\frac{4\sqrt{3}}{f_N}
\int
[\mathrm{d}^2\mathbf{k}_{\perp}]_3
\tilde \psi(\{x_i, \mathbf{k}_{i\perp}\})
\left[\Xi_\uparrow^{p}(\uparrow, \downarrow,\uparrow)+
 \Xi_\uparrow^{p}(\downarrow, \uparrow,\uparrow)\right],\\
\label{A-final}
A&=&
-\frac{4\sqrt{3}}{f_N}
\int
[\mathrm{d}^2\mathbf{k}_{\perp}]_3
\tilde \psi(\{
x_i, \mathbf{k}_{i\perp}\})
\left[\Xi_\uparrow^{p}(\downarrow, \uparrow,\uparrow)-
 \Xi_\uparrow^{p}(\uparrow, \downarrow,\uparrow)\right],\\
\label{T-final}
T&=&
\frac{4\sqrt{3}}{f_N}
\int
[\mathrm{d}^2\mathbf{k}_{\perp}]_3
\tilde \psi(\{
x_i, \mathbf{k}_{i\perp}\})
\Xi_\uparrow^{p}(\uparrow, \uparrow,\downarrow),\\
\label{Phi-final}
\Phi&=&{} -
\frac{8\sqrt{3}}{f_N}
\int
[\mathrm{d}^2\mathbf{k}_{\perp}]_3
\tilde \psi(\{
x_i, \mathbf{k}_{i\perp}\})
\Xi_\uparrow^{p}(\uparrow, \downarrow,\uparrow),
\end{eqnarray}
where the explicit expressions for the $\Xi_\lambda^{p}$ function are given in the Appendix.
These results confirm that by integrating out the transverse-momentum dependence of the nucleon wave function, DAs are determined by its $L_z=0$ component. In particular, in the present SU(6) symmetric model they involve only one of the two independent light-cone amplitudes parametrizing the S-wave component of the LCWF. In order to probe also the other light-cone amplitude one should consider a more general framework with mixed-symmetry terms~\cite{Pasquini:2008}.

In the following the results of Eqs.~(\ref{V-final})-(\ref{Phi-final}) are applied to a specific CQM taking the form of the momentum wave function from Ref.~\cite{Schlumpf:1992ce}, i.e.
\bea
\psi(\{x_i,\boldsymbol{k}_{i\perp}\})=
2(2\pi)^3\bigg[\frac{1}{M_0}\frac{\omega_1\omega_2\omega_3}{x_1x_2x_3}\bigg]^{1/2}
\frac{N'}{(M_0^2+\beta^2)^\gamma},
\label{eq:30}
\eea 
where $\omega_i$ is the free-quark energy and $N'$ is a normalization factor 
such that 
${\int{\rm d}[x]_3\vert \psi(\{x_i\},\{{\tvec k}_{i\perp}\})\vert^2=1}$.
In Eq.~(\ref{eq:30}), the scale $\beta$, 
the parameter $\gamma$ for the power-law behaviour, and the quark mass $m$ 
are taken from Ref.~\cite{Schlumpf:1992ce}, i.e. $\beta=0.607 $ GeV, 
$\gamma=3.4$ and $m=0.267$ GeV. According to the analysis of 
Ref.~\cite{Schlumpf:1992pp} these values lead to a very good description 
of many baryonic properties.

In Fig.~\ref{fig:figure1_DAs} the model results for the proton distribution amplitude $\Phi$ are shown. The resulting shape is quite similar to that of the symmetric asymptotic DA in Eq.~(\ref{Phi_as}). 

\begin{figure}[ht]
\begin{center}
\epsfig{file=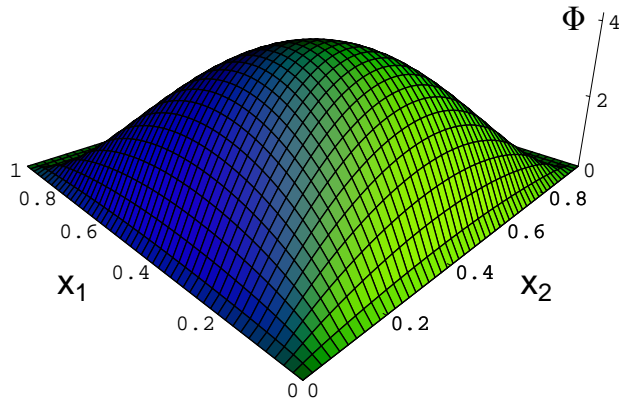,  width=20 pc}
\end{center}
\caption{The proton distribution amplitude.}
\label{fig:figure1_DAs}
\end{figure}

In Table 1 we list the $\phi^{(l,m,n)}$ moments up to the order  $M=l+m+n\leq 3$ in comparison with different model calculations. At this order there are 20 moments out of which only 10 are independent.
Despite the fact that $\Phi$ is normalized, one has to keep in mind that it is a distribution amplitude, not a probability density. Thus its moments cannot strictly be interpreted as mean values~\cite{Stefanis:1987vr}. However,  the first moments of $\Phi$ provide an indication on how the longitudinal momentum of the proton is partitioned among the valence quarks. Since $\phi^{(1,0,0)}\approx 0.6$ in the model of Ref.~\cite{Chernyak:1987nv}, this is the reason why it is claimed that with QCD sum rules about 60\% of the proton longitudinal momentum is carried by the up quark with helicity parallel to that of the proton. 
From Table~\ref{Moments_comparison} we see that a quite different 
result is obtained in the light-cone quark model, a roughly symmetric 
result similar to the one considered in Ref.~\cite{Bolz:1996sw}. In 
fact, while starting from quite different assumptions, the LCWF used 
here and the model wave function of Ref.~\cite{Bolz:1996sw} have some 
common features. Besides the longitudinal momentum dependence of $L_z=0$ 
part of the present LCWF giving rise to an almost symmetric DA, 
our model resembles the Gaussian fall-off as function 
of the quark transverse momentum at 
large $x$ of the wave function of Ref.~\cite{Bolz:1996sw} and 
both models are able, for example, to reproduce the nucleon form factors.
As a matter of fact, the transverse-momentum dependence 
of the present-model 
LCWF has been studied in Ref.~\cite{Boffi:2009sh}, showing that 
the ratio of the squared mean transverse momentum and the mean square 
transverse momentum of transverse momentum dependent parton distributions 
agrees within $10\%$ accuracy with the results obtained assuming a factorized 
gaussian form for the nucleon wave function.

Using these results, we can study the behaviour of proton DA calculated in 
our model after evolution from the initial low-scale of the model to higher 
scales. The initial scale,  corresponding to the results shown in 
Table~\ref{Moments_comparison}, has been fixed evolving back unpolarized data,
 until the valence distributions matches the condition that the second moment,
 i.e. the momentum fraction carried by the valence quarks, is equal to one.
Using LO evolution equations, 
we find $Q^2_0=0.079$ GeV$^2$, with
$\Lambda_{{\rm QCD}} =0.232$ GeV~\cite{Boffi:2003yj}.
Although there is no rigorous relation between the QCD quarks and the 
constituent quarks, and a more fundamental description of the transition from 
soft to hard regimes would be very helpful, this strategy reflects the present 
state of the art for quark model 
calculations~\cite{Scopetta:1997wk,Broniowski:2007si,Pasquini:2004gc},
and has been validated with a fair comparison to 
experiments (see, for example, Ref.~\cite{Boffi:2009sh}). 
The results for the first moments $\phi^{(l,m,n)}$ with $l+m+n\leq 2$ 
after evolution to $Q^2=1$ GeV$^2$ are shown in Table~\ref{Moments_Evol}.
Comparison with the analogous results from lattice estimates~\cite{Gockeler:2008xv} is quite nice. Evolution has only a very small effect: a further evolution to $Q^2=4$ GeV$^2$ would only hardly modify the last digit of our result. This is in agreement with the logarithmic scale dependence predicted by Eq.~(\ref{evolution}) and the fact that already at the input scale the behaviour of our DA approaches that of the asymptotic DA. 

\begin{table}[ht]
\caption{The moments $\phi^{(l,m,n)}$ of the proton DA at different scales.
The asymptotic values (AS) of Eq.~(\ref{Phi_as}) (second column)
are compared with the lattice results (LAT)~\cite{Lenz:2009ar} at the scale $Q^2=1$ GeV$^2$ (third column) and with the corresponding results from the present model calculation (PPB), after evolution from the initial scale $Q^2_0=0.079$ GeV$^2$ to $Q^2=1$ GeV$^2$ (last column).}
\label{Moments_Evol}
\begin{center}
\begin{tabular}{ccll}
\hline\hline\\
($l$,\,$m$,\,$n$)   
&
AS
&
\qquad LAT
&
PPB\\ \\
\hline\\
$(0\;\; 0\;\; 0)\,$ & $\,    1    \,$                        & $\,   1    \,$ & $\,   1    \,$ \\
$(1\;\; 0\;\; 0)\,$ & $\,\onet\simeq 0.333\,$ & $\,     0.3999(37)(139)   \,$  &  $\, 0.340\,$  \\
$(0\;\; 1\;\; 0)\,$ & $ \onet\simeq 0.333$ & $\,   0.2986(11)(52)    \,$ &  $\, 0.335\,$ \\
$(0\;\; 0\;\; 1)\,$ & $\onet\simeq 0.333$& $\, 0.3015(32)(106) \,$ &$\, 0.326\,$\\
$(2\;\; 0\;\; 0)\,$ & $\oneseven\simeq 0.143$ & $\, 0.1816(64)(212) \,$ &$\, 0.147\,$\\
$(0\;\; 2\;\; 0)\,$ & $\oneseven\simeq 0.143$ & $\,  0.1281(32)(106)  \,$ &$\, 0.144\,$\\
$(0\;\; 0\;\; 2)\,$ & $\oneseven\simeq 0.143$& $\,  0.1311(113)(382)  \,$ &$\,0.137\,$ \\
$(1\;\; 1\;\; 0)\,$ & ${\textstyle {2\over 21}}\simeq 0.095$ & $\, 0.1092(67)(219)  \,$ &$\, 0.098\,$\\
$(1\;\; 0\;\; 1)\,$ & ${\textstyle {2\over 21}}\simeq 0.095$ & $\,  0.1091(41)(152)  \,$ &$\, 0.095\,$\\
$(0\;\; 1\;\; 1)\,$ & ${\textstyle {2\over 21}}\simeq 0.095$ & $\,  0.0613(89)(319)  \,$ &$\,0.093\,$ \\
\\
\hline\hline
\end{tabular}
\end{center}
\end{table}


\section{Transition distribution amplitudes in a meson-cloud model}
\label{Sec:TDAs}

The general matrix element describing the transition from a nucleon
to a meson state reads~\cite{Pire:2005ax}
\begin{equation}\label{General-Matrix-Element-TDA}
\langle\pi|\epsilon^{ijk}q^{i'}_\alpha(z_1n)[z_1;z_0]_{i'i}
q^{j'}j_\beta(z_2n)[z_2;z_0]_{j'j}
q^{k'}_\gamma(z_3n)[z_3;z_0]_{k'k}|N\rangle,
\end{equation}
where the Wilson lines  $[z_i ; z_0]$ are defined as in Eq.~(\ref{wilson}).
In the following they will be neglected by assuming to work in the light-cone gauge $A^+=0$.
The spinorial and Lorentz decomposition of the matrix
element~(\ref{General-Matrix-Element-TDA}) follows the same line as in
the case of the baryon DAs.
In particular, for the $p\rightarrow \pi^0$ transition the leading-twist TDAs can be defined as
\begin{eqnarray}
&&4\mathcal{F}\bigg(\langle\pi^0(p_\pi)|\epsilon^{ijk}u^i_\alpha(z_1n)
u^j_\beta(z_2n) d^k_\gamma(z_3n)|p(p_p,\lambda)\rangle\bigg) 
\nonumber\\
&&{}= i\frac{f_N}{f_{\pi}}\Big[V_1^{p\pi^0}(\slashed{p}
C)_{\alpha\beta}(N^+)_\gamma + A_1^{p\pi^0}(\slashed{p}
\gamma^5C)_{\alpha\beta}(\gamma^5N^+)_\gamma\nonumber\\ 
&&{}\quad +
T_1^{p\pi^0}(\sigma_{p\mu}C)_{\alpha\beta}(\gamma^\mu N^+)_\gamma +
M^{-1}V_2^{p\pi^0}(\slashed{p}C)_{\alpha\beta}(\slashed{\Delta}_\perp N^+)_\gamma\nonumber\\
&&{}\quad+
M^{-1}A_2^{p\pi^0}(\slashed{p}\gamma^5C)_{\alpha\beta}(\gamma^5\slashed{\Delta}_\perp N^+)_\gamma
+ M^{-1}T_2^{p\pi^0}(\sigma_{p\Delta_\perp}
C)_{\alpha\beta}(N^+)_\gamma\nonumber\\
&&{}\quad+ M^{-1}T_3^{p\pi^0}(\sigma_{p\mu}
C)_{\alpha\beta}(\sigma^{\mu\Delta_\perp}N^+)_\gamma +
M^{-2}T_4^{p\pi^0}(\sigma_{p\Delta_\perp}
C)_{\alpha\beta}(\slashed{\Delta}_\perp N^+)_\gamma\Big],
\label{decomposition}
\end{eqnarray}
where the symbol $\mathcal{F}$ represents the Fourier transform (like Eq.~(\ref{eq:fourier})) and
$f_\pi$ is the pion decay constant ($f_\pi= \sqrt{2}F_\pi = 131$ MeV).
In a reference frame with the $z$-axis along the direction of 
the proton momentum, the pion momentum $p_\pi$ and the proton momentum $p_p$ have the following Sudakov decomposition
\begin{eqnarray}
&& p_p = (1 + \xi)p + \frac{M^2}{1 + \xi}n,  \\
&& 
p_\pi = (1 - \xi)p + \frac{m_\pi^2 +  \Delta^2_\perp}{1 - \xi}n +
\boldsymbol{\Delta}_\perp,
\eea
where $\Delta$ is the  four-momentum transfer,
\bea
&& \Delta = p_\pi - p_p = -2\xi p + \bigg[\frac{m_\pi^2 +
\Delta^2_\perp}{1 - \xi} - \frac{M^2}{1 + \xi}\bigg]n + 
\boldsymbol{\Delta_\perp},
\label{eq:kin}
\end{eqnarray}
and  $\xi$
is the skewness variable describing the loss of plus momentum of the initial hadron in the proton-to-meson transition, i.e.
\be
\xi = -\frac{\Delta\cdot n}{2P\cdot n}=-\frac{\Delta^+}{2P^+}, \quad
\mbox{with}\;\;
P =\frac{1}{2}(p_p+ p_\pi).
\ee
The TDAs are dimensionless functions and depend on $(x_i,\xi,\Delta^2)$,
where the fraction of quark plus momentum $x_i$ have support in $[-1+\xi,1+\xi]$ and 
\be
\Delta^2= -2\xi\left[\frac{m_\pi^2+\Delta_\perp^2}{1-\xi}
-\frac{M^2}{1+\xi}\right]-\Delta_\perp^2.
\ee
Restricting ourselves 
to the case $\xi>0$, momentum conservation requires $\sum_ix_i=2\xi$.
The fields with positive momentum fractions, $x_i\geq 0$, describe creation of quarks, whereas those with negative fractions, $x_i\leq 0$, the absorption of antiquarks.
This leads to define three distinct kinematical regions:
the ERBL region for $x_i\geq 0$, and two DGLAP regions when 
 $x_1\geq 0,$ $x_2\geq 0,$ $x_3\leq 0,$  or $x_1\geq 0,$ $x_2\leq 0,$ $x_3\leq 0.$ 
The names derive from the evolution equations which controls the scale 
dependence of the TDAs in the different regions.

Denoting the matrix element in left-hand side of Eq.~(\ref{decomposition}) 
by $T_{\alpha\beta,\gamma}^{\lambda}$, we can derive the eight TDAs in terms of
the following linear combinations of matrix elements
\bea
\label{V1pion}
  V_1^{p\pi_0} &=& -i\frac{1}{2^{1/4}\sqrt{1+\xi}(P^+)^{3/2}}\frac{f_\pi}{f_N}\Big(T_{12,1}^\uparrow + T_{21,1}^\uparrow\Big),
  \\
\label{A1pion}
  A_1^{p\pi_0} &=& i\frac{1}{2^{1/4}\sqrt{1+\xi}(P^+)^{3/2}}\frac{f_\pi}{f_N}
\Big(T_{12,1}^\uparrow-T_{21,1}^\uparrow  \Big),
  \\
\label{T1pion}
  T_1^{p\pi_0} &=& i\frac{1}{2^{1/4}\sqrt{1+\xi}(P^+)^{3/2}}\frac{f_\pi}{f_N}
\bigg[T_{11,2}^\uparrow
  + \frac{(\Delta^-_\perp)^2}{\Delta_{\perp}^2}T_{22,2}^\uparrow\bigg],
\\
\label{V2pion}
  V_2^{p\pi_0} &=& -i\frac{M\Delta^-_\perp}{\Delta_{\perp}^{2}}
\frac{1}{2^{1/4}\sqrt{1+\xi}(P^+)^{3/2}}\frac{f_\pi}{f_N}
\Big(T_{12,2}^\uparrow
+ T_{21,2}^\uparrow\Big),
\\
\label{A2pion}
  A_2^{p\pi_0}&=& -i\frac{M\Delta^-_\perp}{\Delta_{\perp}^2}
\frac{1}{2^{1/4}\sqrt{1+\xi}(P^+)^{3/2}}\frac{f_\pi}{f_N}
\Big(T_{12,2}^\uparrow
- T_{21,2}^\uparrow\Big),
\\
 \label{T2pion}
  T_2^{p\pi_0} &=& -i\frac{M}{\Delta_{\perp}^2}
\frac{1}{2^{1/4}\sqrt{1+\xi}(P^+)^{3/2}}\frac{f_\pi}{f_N}
\left[\Delta^+_\perp\,T_{11,1}^\uparrow
- \Delta^-_\perp\,T_{22,1}^\uparrow\right],
\\
\label{T3pion}
  T_3^{p\pi_0} &=&i\frac{M}{\Delta_{\perp}^2}
\frac{1}{2^{1/4}\sqrt{1+\xi}(P^+)^{3/2}}\frac{f_\pi}{f_N}
\left[\Delta^+_\perp\,T_{11,1}^\uparrow
+  \Delta^-_\perp\,T_{22,1}^\uparrow\right],\\
\label{T4pion}
  T_4^{p\pi_0} &=& i\frac{2M^2(\Delta^-_\perp)^2}{(\Delta_{\perp}^2)^2}
\frac{1}{2^{1/4}\sqrt{1+\xi}(P^+)^{3/2}}\frac{f_\pi}{f_N}
T_{22,2}^\uparrow,
  \eea
where $\Delta^\pm_\perp=\Delta_x\pm i\Delta_y$.

In the following we focus on the study of the TDAs in the ERBL region, 
corresponding to probe the $\psi_{3q+q\bar q}$ 
Fock-component of the nucleon wave function.
The five-parton component of the nucleon state can be modeled using the meson-cloud model developed in Refs.~\cite{Pasquini:2006dv,Pasquini:2007iz}.
The basic assumption of the model 
is that the physical nucleon is made of a bare nucleon 
dressed by the surrounding meson cloud so that  the nucleon state is decomposed according to the meson-baryon Fock-state expansion as a superposition of a bare  nucleon, formed by three valence quarks, and states containing virtual mesons 
with recoiling baryons.
These baryon-meson subsystems are assumed to include configurations with the baryon being a nucleon or a $\Delta$ and the accompanying meson being a pion as well as a vector meson such as the $\rho$ or the $\omega$.
Being interested to the $p\rightarrow \pi^0$ TDAs, here we will consider the 
meson-baryon components with a pion and a nucleon or a $\Delta$, given by the following representation in the light-cone dynamics
\begin{eqnarray} 
|N(B\pi); p_p, \lambda\rangle &=&
\int\mathrm{d}y\mathrm{d}^2\mathbf{k}_\perp\int_0^y\prod_{i=1}^3\frac{\mathrm{d}\xi_i}{\sqrt{\xi_i}}
\int_0^{(1-y)}\prod_{i=4}^{5}\frac{\mathrm{d}\xi_i}{\sqrt{\xi_i}}
\int
\frac{1}{[2(2\pi)^3]^4}
\prod_{i=1}^{5}\mathrm{d}^2\mathbf{k}'_{i\perp}\nonumber\\\nonumber
&&\times\delta\bigg(y - \sum_{i=1}^3
\xi_i\bigg)\delta^{(2)}\bigg(\mathbf{k}_{\perp} -
\sum_{i={1}}^{3}\mathbf{k}'_{i\perp}\bigg)\delta\bigg(1 -
\sum_{i={1}}^{5}\xi_i\bigg)\delta^{(2)}\bigg(\sum_{i={1}}^{5}\mathbf{k}'_{i\perp}\bigg)\nonumber\\\nonumber
&&\times \sum_{\lambda'}\sum_{\lambda_i, \tau_i, c_i}
\phi_{\lambda' 0}^{\lambda(N,B\pi)}(y,\textbf{k}_\perp)
\sqrt{y(1-y)}
\tilde{\Psi}_{\lambda'}^{B,[f]}(y,
\mathbf{k}_\perp; \{\xi_i, \mathbf{k}'_{i\perp}; \lambda_i, \tau_i,
c_i\}_{i=1,\cdots,3})\nonumber\\
&&\times\tilde{\Psi}^{\pi,[f]}(1-y,-\mathbf{k}_\perp;\{\xi_i,
\mathbf{k}'_{i\perp}; \lambda_i,
\tau_i\}_{i=4,5})\prod_{i={1}}^{5}|\xi_ip_p^+,
\mathbf{k}'_{i\perp} + \xi_i\mathbf{p}_{p\perp}, \lambda_i, \tau_i,
c_i; q\rangle,
\label{meson-cloud}
\end{eqnarray}
where the LCWF of the baryon,
$\tilde\Psi^{B,[f]}_{\lambda'}$, and  the pion,
$\tilde\Psi^{\pi,[f]}$, 
incorporate the Jacobian ${\cal J}$ of the transformation from the intrinsic 
variables with respect to the hadron rest-frame 
($\{ \zeta_i,\boldsymbol{\kappa}_{i\perp}\}$) to the intrinsic variables with 
respect to the nucleon rest frame ($\{\xi_i, \mathbf{k}'_{i\perp}\}$), i.e.
\begin{eqnarray}
\tilde \Psi_{\lambda'}^{B,\,[f]}(\{\xi_i,{\bf k}'_{i\perp};\lambda_i,\tau_i\}_{i=1,2,3})
&=&\sqrt{{\cal J}(\xi_1,\xi_2,\xi_3)}
\tilde\Psi_{\lambda'}^{B,\,[f]}(\{\zeta_i,\boldsymbol{\kappa}_{i\perp};\lambda_i,\tau_i\}_{i=1,2,3})
\nn\\
&=&
\frac{1}{y^{3/2}}
\tilde\Psi_{\lambda'}^{B,\,[f]}(\{\zeta_i,\boldsymbol{\kappa}_{i\perp}
;\lambda_i,\tau_i\}_{i=1,2,3}),
\label{eq:3q}
\end{eqnarray}
\begin{eqnarray}
\tilde \Psi^{\pi,\,[f]}(\{\xi_i,{\bf k}'_{i\perp};\lambda_i,\tau_i\}_{i=4,5})
&=&\sqrt{{\cal J}(\xi_4,\xi_5)}
\tilde\Psi^{\pi,\,[f]}(\{\zeta_i,\boldsymbol{\kappa}_{i\perp};\lambda_i,\tau_i\}_{i=4,5})
\nn\\
&=&\frac{1}{(1-y)}
\tilde\Psi^{\pi,\,[f]}(\{\zeta_i,\boldsymbol{\kappa}_{i\perp};\lambda_i,\tau_i\}_{i=4,5}).
\label{eq:5q}
\end{eqnarray}
The relations between the variables ($\{x_i, \mathbf{k}_{i\perp} \}$)
and $\{\xi_i, \mathbf{k}'_{i\perp}\}$, are given by:

For $ i=1,2,3$, corresponding to the indices of the three quarks in the baryon,
\begin{eqnarray}
\zeta_i =\frac{\xi_i}{y},\qquad
\boldsymbol{\kappa}_{i\perp}={\bf k}'_{i\perp}-\zeta_i\,{\bf k}_\perp;
\label{baryon_var}
\end{eqnarray}

For $i=4,5$, corresponding to the indices of the quark and antiquark in the pion, respectively:
\begin{eqnarray}
\zeta_i =\frac{\xi_i}{(1-y)},\qquad
\boldsymbol{\kappa}_{i\perp}={\bf k}'_{i\perp}+\zeta_i\,{\bf k}_\perp.
\label{meson_var}
\end{eqnarray}

In Eq.~(\ref{meson-cloud})
the function 
$\phi_{\lambda'0}^{\lambda\,(N,B\pi)}(y,{\mathbf k}_\perp)$ is the probability amplitude to find a physical nucleon with helicity $\lambda$
in a state consisting of a  virtual baryon $B=N,\Delta$ 
and a virtual pion,
 with the baryon having helicity $\lambda'$, longitudinal 
momentum fraction $y$ and transverse momentum ${\mathbf k}_\perp$, and the 
pion having
longitudinal momentum fraction $1-y$ and 
transverse momentum $
-{\mathbf k}_\perp$.
This splitting function can be calculated using time-ordered perturbation 
theory in the infinite momentum frame as explained in Ref.~\cite{Speth:1998}, and 
have also been rederived and tabulated in Ref.~\cite{Pasquini:2006dv}.

For the pion state in the matrix element of Eq.~(\ref{decomposition})
we consider the valence $q\bar q$ component given by
\be
\label{pion-state}
\ket{\pi(p_\pi)} = \sum_{\tau_i,\lambda_i,c_i}
\int\left[\frac{{\rm d}\xi }{\sqrt{\xi}}\right]_2
[{\rm d}^2{\bf k}_\perp]_2
\Psi^{\pi,[f]}(\{\xi_i,{\bf k}_{i\perp};\lambda_i,\tau_i\}_{i=1,2})
\prod_{i=1}^2
\ket{q^{\lambda_i};\xi_i p^+_\pi, \, {\bf p}_{i\perp}},
\ee
where ${\bf p}_{\perp i}={\bf k}_{i\perp}+\xi_ip^+_\pi$ and the $q\bar q$ state is defined as
\be
\prod_{i=1}^2
\ket{q^{\lambda_i};\tilde k_i}=
\frac{\delta_{ij}}{\sqrt{3}}
b_q^{\dagger i}(\tilde k_1,\lambda_1)d_q^{\dagger j}(\tilde k_2,\lambda_2)|0\rangle.
\ee

Using the expressions for the proton and pion state given in Eqs.~(\ref{meson-cloud}) and ~(\ref{pion-state}), and 
 the momentum-space expansion (\ref{Free-quark-field}) of the quark fields, 
combined with the anticommutation relations for the quark creation 
and annihilation operators, 
the final expression for the matrix elements 
$T^{\lambda}_{\alpha,\beta\gamma}$ is given by
\begin{eqnarray}
T^{\lambda}_{\alpha,\beta\gamma}=
&&{}-
24 \left(\frac{1}{2\xi}\right)^{3/2}
\frac{1}{\sqrt{x_1x_2x_3}}
u_{+\alpha}(k^+_1,
\lambda_1)u_{+\beta}(k^+_2, \lambda_2)u_{+\gamma}(k^+_3,
\lambda_3)\nonumber\\
&&{}\times
\sum_B
\sum_{\lambda'}
\int\mathrm{d}y\mathrm{d}^2\mathbf{k}_\perp
\phi_{\lambda' 0}^{\lambda (N, B\pi)}(y,\textbf{k}_\perp)\sqrt{y(1-y)}
\delta\bigg(1 -y-\frac{p_\pi^+}{p_p^+}\bigg)
\delta^{(2)}\bigg(\mathbf{k}_{\perp} +
\mathbf{p}_{\pi\perp}
\bigg)
\nonumber\\
&&{}\times
\int
[\mathrm{d}^2\boldsymbol{\kappa}_{\perp}]_3\,
\tilde{\Psi}_{\lambda'}^{B,[f]}\bigg(
\{\frac{x_1}{2\xi}, \boldsymbol{\kappa}_{1\perp}; \lambda_{1}, 1/2\}\{\frac{x_2}{2\xi}, \boldsymbol{\kappa}_{2\perp}; \lambda_{2}, 1/2\}\{\frac{x_3}{2\xi},
\boldsymbol{\kappa}_{3\perp}; \lambda_{3},
-1/2\}\bigg).\nonumber\\& &
\label{eq_pion6}
\end{eqnarray}
The light-cone spinors of the quarks are defined as in Eq.~(\ref{LC-spinors}), and depend on the longitudinal momenta $k^+_i=x_iP^+$.
The Dirac indices $\alpha,$ $\beta,$ and $\gamma$ fix the total quark helicity 
of the baryon wave function, as explained in sect.~\ref{sec:DA}, while
the isospin quantum numbers in the baryon wave function correspond to the $uud$ configuration.
Eq.~(\ref{eq_pion6}) has a clear physical interpretation
and allows to relate the nucleon-to-pion TDAs
to the baryon distribution amplitudes in the $(B\pi)$ component of the nucleon
 weighted by the probability amplitude that the nucleon fluctuates
in the corresponding $(B\pi)$ subsytem with the pion momentum matching 
the pion momentum in the final state.
The momentum fraction of the quarks in the baryon LCWF are
defined with respect to the longitudinal momentum of the baryon, 
i.e. $\kappa^+_i/(yp^+_p)=x_i/(2\xi)$, while the integration over the transverse
quark momenta corresponds to the projection of the baryon LCWF onto the zero orbital angular momentum component.
In Eq.~(\ref{eq_pion6}), the sum over the baryon states is 
restricted to the nucleon and the $\Delta$, while the sum over the helicity $\lambda'$ 
of the baryon permits baryon-pion fluctuations  which
do not conserve the helicity of the parent nucleon.

In the case of the nucleon contribution,
we model the proton LCWF as in Eq.~(\ref{LCWFuud}), with parameters 
$\gamma=3.21$, $\beta=0.489$ GeV and $m=0.264$ GeV from the fit of the valence
 and meson-cloud contribution to  the electroweak nucleon form factors~\cite{Pasquini:2007iz}.
The $\Delta$ is described as a state of isospin $\tau=\treh$ obtained as a pure splin-flip excitation of the nucleon, with the same momentum-dependent wave function of the nucleon, i.e.
\begin{eqnarray}
\label{LCWFuud-delta}
\Psi^{\Delta,[f]}_\lambda(\{x_i\}, \{k_i\}, \{\lambda_i\}, \{uud\}) 
&=& \tilde \psi(\{x_i,{\bf k}_{i\perp}\})
\Xi_{\lambda}^{\Delta}(\lambda_1,\lambda_2,\lambda_3)
\tilde{\Phi}^{1}_{\treh\oneh}(\oneh,\oneh,-\oneh),
\eea
with the isospin coefficient $\tilde{\Phi}^{1}_{\treh\oneh}(\frac{1}{2},\frac{1}{2},-\frac{1}{2} )
=\sqrt{\frac{1}{3}}$ and the spin dependent part  given by
\begin{eqnarray}
\Xi_{\lambda}^{\Delta}(\lambda_1,\lambda_2,\lambda_3)=\sum_{\mu_1\mu_2\mu_3}
D_{\mu_1\lambda_1}^{1/2*}(R_{cf}(\tilde k_1))
D_{\mu_2\lambda_2}^{1/2*}(R_{cf}(\tilde k_2))
D_{\mu_3\lambda_3}^{1/2*}(R_{cf}(\tilde k_3))
\tilde{\Phi}^{1}_{\treh\lambda}(\mu_1, \mu_2,\mu_3).
\end{eqnarray}
The explicit expression of the functions $\Xi_{\lambda}^{\Delta}$ for all the possible spin configurations 
of the three quarks in the $\Delta$ state is given in the Appendix.
\newline
\noindent
Finally, the splitting function $\phi^{(N,B\pi)}_{\lambda 0}$ in 
Eq.~(\ref{eq_pion6}) is calculated as in Ref.~\cite{Pasquini:2007iz}.

Inserting the matrix elements $T^{\lambda}_{\alpha,\beta\gamma}$ from Eq.~(\ref{eq_pion6}) into Eqs.~(\ref{V1pion})-(\ref{T4pion}) we finally obtain the expressions of the TDAs in the meson-cloud model:

\bea
  V_1^{p\pi_0} &=& i\frac{4\sqrt{3}}{2\xi}\sqrt{\frac{(1-\xi)}{(1+\xi)^3}}
\frac{f_\pi}{f_N}
\int
[\mathrm{d}^2\mathbf{k}_{\perp}]_3
\tilde \psi(\{x_i, \mathbf{k}_{i\perp}\})
\nonumber\\
&&\times\left\{
\phi_{\uparrow 0}^{\uparrow (N, N\pi)}(y
,-\textbf{p}_{\pi\perp})
\left[
\Xi^N_\uparrow(\uparrow,\downarrow,\uparrow)
+\Xi^N_\uparrow(\downarrow,\uparrow,\uparrow)\right]
+
\phi_{\uparrow 0}^{\uparrow (N, \Delta\pi)}(y
,-\textbf{p}_{\pi\perp})
\left[
\Xi^\Delta_\uparrow(\uparrow,\downarrow,\uparrow)+\Xi^\Delta_\uparrow(\downarrow,\uparrow,\uparrow)\right]\right\},
\label{V1pion-mc}
\\
  A_1^{p\pi_0} &=& -i\frac{4\sqrt{3}}{2\xi}\sqrt{\frac{(1-\xi)}{(1+\xi)^3}}
\frac{f_\pi}{f_N}
\int
[\mathrm{d}^2\mathbf{k}_{\perp}]_3
\tilde \psi(\{x_i, \mathbf{k}_{i\perp}\})
\nonumber\\
&&\times\left\{
\phi_{\uparrow 0}^{\uparrow (N, N\pi)}(y
,-\textbf{p}_{\pi\perp})
\left[
\Xi^N_\uparrow(\uparrow,\downarrow,\uparrow)
-\Xi^N_\uparrow(\downarrow,\uparrow,\uparrow)\right]
+
\phi_{\uparrow 0}^{\uparrow (N, \Delta\pi)}(y
,-\textbf{p}_{\pi\perp})
\left[
\Xi^\Delta_\uparrow(\uparrow,\downarrow,\uparrow)-\Xi^\Delta_\uparrow(\downarrow,\uparrow,\uparrow)\right]\right\},
\label{A1pion-mc}
\\
  T_1^{p\pi_0} &=& -i\frac{4\sqrt{3}}{2\xi}\sqrt{\frac{(1-\xi)}{(1+\xi)^3}}
\frac{f_\pi}{f_N}
\int
[\mathrm{d}^2\mathbf{k}_{\perp}]_3
\tilde \psi(\{x_i, \mathbf{k}_{i\perp}\})\left\{
\phi_{\uparrow 0}^{\uparrow (N, N\pi)}(y,-\textbf{p}_{\pi\perp})
\Xi^N_\uparrow(\uparrow,\uparrow,\downarrow)\right.\nonumber\\
&&
\left.
+
\phi_{\uparrow 0}^{\uparrow (N, \Delta\pi)}(y,-\textbf{p}_{\pi\perp})
\Xi^\Delta_\uparrow(\uparrow,\uparrow,\downarrow)+
\frac{(\Delta^-_\perp)^2}{\Delta_{\perp}^2}
\phi_{\Downarrow 0}^{\uparrow (N, \Delta\pi)}(y,-\textbf{p}_{\pi\perp})
\Xi^\Delta_\Downarrow(\downarrow,\downarrow,\downarrow)\right\},
\label{T1pion-mc}
\\
  V_2^{p\pi_0} &=&i\frac{4\sqrt{3}}{2\xi}\sqrt{\frac{(1-\xi)}{(1+\xi)^3}}
\frac{f_\pi}{f_N}
\frac{M\Delta^-_\perp}{\Delta_{\perp}^{2}}
\int
[\mathrm{d}^2\mathbf{k}_{\perp}]_3
\tilde \psi(\{x_i, \mathbf{k}_{i\perp}\})
\nonumber\\
&&\times\left\{
\phi_{\downarrow 0}^{\uparrow (N, N\pi)}(y,-\textbf{p}_{\pi\perp})
\left[
\Xi^N_\downarrow(\uparrow,\downarrow,\downarrow)
+
\Xi^N_\downarrow(\downarrow,\uparrow,\downarrow)\right]
+
\phi_{\downarrow 0}^{\uparrow (N, \Delta\pi)}(y,-\textbf{p}_{\pi\perp})
\left[
\Xi^\Delta_\downarrow(\uparrow,\downarrow,\downarrow)
+
\Xi^\Delta_\downarrow(\downarrow,\uparrow,\downarrow)\right]
\right\},
\label{V2pion-mc}
\\
  A_2^{p\pi_0}&=& i\frac{4\sqrt{3}}{2\xi}\sqrt{\frac{(1-\xi)}{(1+\xi)^3}}
\frac{f_\pi}{f_N}
\frac{M\Delta^-_\perp}{\Delta_{\perp}^{2}}
\int
[\mathrm{d}^2\mathbf{k}_{\perp}]_3
\tilde \psi(\{x_i, \mathbf{k}_{i\perp}\})
\nonumber\\
&&\times\left\{
\phi_{\downarrow 0}^{\uparrow (N, N\pi)}(y,-\textbf{p}_{\pi\perp})
\left[
\Xi^N_\downarrow(\uparrow,\downarrow,\downarrow)
-
\Xi^N_\downarrow(\downarrow,\uparrow,\downarrow)\right]
+
\phi_{\downarrow 0}^{\uparrow (N, \Delta\pi)}(y,-\textbf{p}_{\pi\perp})
\left[
\Xi^\Delta_\downarrow(\uparrow,\downarrow,\downarrow)
-
\Xi^\Delta_\downarrow(\downarrow,\uparrow,\downarrow)\right]
\right\},
\label{A2pion-mc}
\\
  T_2^{p\pi_0} &=& i\frac{4\sqrt{3}}{2\xi}\sqrt{\frac{(1-\xi)}{(1+\xi)^3}}
\frac{f_\pi}{f_N}
\int
[\mathrm{d}^2\mathbf{k}_{\perp}]_3
\tilde \psi(\{x_i, \mathbf{k}_{i\perp}\})\left\{
- \Delta^-_\perp\,
\phi_{\downarrow 0}^{\uparrow (N, N\pi)}(y,-\textbf{p}_{\pi\perp})
\Xi^N_\downarrow(\downarrow,\downarrow,\uparrow)\right.\nonumber\\
&&
\left.
+\Delta^+_\perp\
\phi_{\Uparrow 0}^{\uparrow (N, \Delta\pi)}(y,-\textbf{p}_{\pi\perp})
\Xi^\Delta_\Uparrow(\uparrow,\uparrow,\uparrow)
- \Delta^-_\perp\,
\phi_{\downarrow 0}^{\uparrow (N, \Delta\pi)}(y,-\textbf{p}_{\pi\perp})
\Xi^\Delta_\downarrow(\downarrow,\downarrow,\uparrow)
\right\},
 \label{T2pion-mc}
\\
  T_3^{p\pi_0} &=& -i\frac{4\sqrt{3}}{2\xi}\sqrt{\frac{(1-\xi)}{(1+\xi)^3}}
\frac{f_\pi}{f_N}
\int
[\mathrm{d}^2\mathbf{k}_{\perp}]_3
\tilde \psi(\{x_i, \mathbf{k}_{i\perp}\})\left\{
\Delta^-_\perp\,
\phi_{\downarrow 0}^{\uparrow (N, N\pi)}(y,-\textbf{p}_{\pi\perp})
\Xi^N_\downarrow(\downarrow,\downarrow,\uparrow)\right.\nonumber\\
&&
\left.
+\Delta^+_\perp\
\phi_{\Uparrow 0}^{\uparrow (N, \Delta\pi)}(y,-\textbf{p}_{\pi\perp})
\Xi^\Delta_\Uparrow(\uparrow,\uparrow,\uparrow)
+ \Delta^-_\perp\,
\phi_{\downarrow 0}^{\uparrow (N, \Delta\pi)}(y,-\textbf{p}_{\pi\perp})
\Xi^\Delta_\downarrow(\downarrow,\downarrow,\uparrow)
\right\},
\label{T3pion-mc}
\\
  T_4^{p\pi_0} &=& -i\frac{4\sqrt{3}}{2\xi}\sqrt{\frac{(1-\xi)}{(1+\xi)^3}}
\frac{f_\pi}{f_N}\frac{2M^2(\Delta^-_\perp)^2}{(\Delta_{\perp}^2)^2}
\int
[\mathrm{d}^2\mathbf{k}_{\perp}]_3
\tilde \psi(\{x_i, \mathbf{k}_{i\perp}\})
\phi_{\Downarrow 0}^{\uparrow (N, \Delta\pi)}(y,-\textbf{p}_{\pi\perp})
\Xi^\Delta_\Downarrow(\downarrow,\downarrow,\downarrow),
\label{T4pion-mc}
  \eea
where the longitudinal momentum fraction in the argument of the splitting function 
is  $y=2\xi/(1+\xi)$ and the spin $\treh$ ($-\treh$) state of the $\Delta$ is indicated as $\Uparrow$ ($\Downarrow$).


\section{Some results for the TDAs}
\label{sec:results}

As an example results are reported in 
Figs.~\ref{Fig:V1-A1-V2-A2-p_pi} and~\ref{Fig:T1-T2-T3-T4-p_pi} under kinematic 
conditions relevant in the case of hard exclusive electroproduction of a pion 
in the backward region~\cite{Lansberg:2007ec} or in associated production of a 
pion and a high-$Q^2$ dilepton pair in $p\bar p$ 
annihilation~\cite{Lansberg:2007se}. Preliminary results for other kinematics have been presented in Ref.~\cite{Pincetti:2008fh}.

 The vector and axial-vector TDAs in Fig.~\ref{Fig:V1-A1-V2-A2-p_pi} 
exhibit the expected symmetric and antisymmetric behaviour 
under permutation of the two up quarks, respectively. 
The $\Delta$ contribution  has the same shape as the proton contribution, 
with the same sign for $A^{p\pi}_1$ and $V^{p\pi}_2$, and opposite sign for 
$V^{p\pi}_1$ and $A^{p\pi}_2$.
The relative contribution of the nucleon with respect to the $\Delta$ is 
always smaller in absolute value for the vector TDAs, 
being suppressed by a factor of about 10 in the case of  $V^{p\pi}_1$
and by a factor of about 1.5 in the case of $V^{p\pi}_2$.
Viceversa, for the axial-vector TDAs one finds that the nucleon contribution 
to $A^{p\pi}_1$ is smaller than the $\Delta$ contribution by a factor 3, while
for $A^{p\pi}_2$ the weight of the nucleon contribution is three times larger 
than in the case of the $\Delta.$
This can be traced back both to the different spin structure 
and to the different splitting functions in 
Eqs.~(\ref{V1pion-mc})-(\ref{A1pion-mc}), and 
(\ref{V2pion-mc})-(\ref{A2pion-mc}).
In particular, $V^{p\pi}_1$ and  $A^{p\pi}_1$ involve splitting functions 
without flip of the helicity of the parent nucleon, 
while $V^{p\pi}_2$ and  $A^{p\pi}_2$ are proportional to splitting functions 
with helicity flip.
In the explored kinematics, the $(N,N\pi)$ vertex without helicity flip
is suppressed by a factor 5 with respect to the $(N,\Delta\pi)$ interaction, 
while for the opposite case with helicity flip the probability amplitude 
to have a $(N\pi)$ fluctuation in the nucleon is almost twice larger than
for the $(\Delta\pi)$ subsystem.

The tensor TDAs in Fig.~\ref{Fig:T1-T2-T3-T4-p_pi} are symmetric 
 under permutation of the two up quarks and, at variance with the other TDAs, 
involve  spin configurations also with parallel helicities of all three quarks.
Such spin configurations receive contribution only from the $\Delta$
with helicity $J_z=\pm 3/2$ because of the projection of the baryon wave function in Eqs.~(\ref{T1pion-mc}) and (\ref{T2pion-mc})-(\ref{T4pion-mc}) 
onto the zero orbital angular momentum component.
In particular, one finds that in the explored kinematics the splitting function
of the nucleon into a $\Delta$ with helicity $J_z=3/2$ is 10 times bigger than 
in the case with helicity $J_z=-3/2$, and with opposite sign.
Furthermore, these terms are multiplied by 
kinematical coefficients which modulate their relative contribution 
to the tensor TDAs in a quite different way.
For example, the  contribution of the
$\Delta$ state with helicity $J_z=3/2$ is quite small  in the case 
of $T^{p\pi}_1$, while  it is enhanced by an additional 
factor of $2M^2/\Delta^2_\perp\simeq 50$ in $T^{p\pi}_4.$
This is the only non-vanishing contribution to $T^{p\pi}_4$.
In $T^{p\pi}_1$ there is also the contribution from the 
helicity states $J_z=1/2,$ with the same sign
for the nucleon and the $\Delta$, but larger by about a factor 10 in 
the case of $\Delta$. 
In the case of $T^{p\pi}_2$ and $T^{p\pi}_3$ the contribution from the 
$\Delta$ is large, and it is mainly given by the configuration with helicity $J_z=3/2$. However,  this contribution is reduced by that of the proton with a similar shape in the case of $T^{p\pi}_2$,  whereas 2/3 of $T^{p\pi}_3$ are due to the proton and 1/3 to the $\Delta$.


\begin{figure}[h!]
 	\centering
\epsfig{file=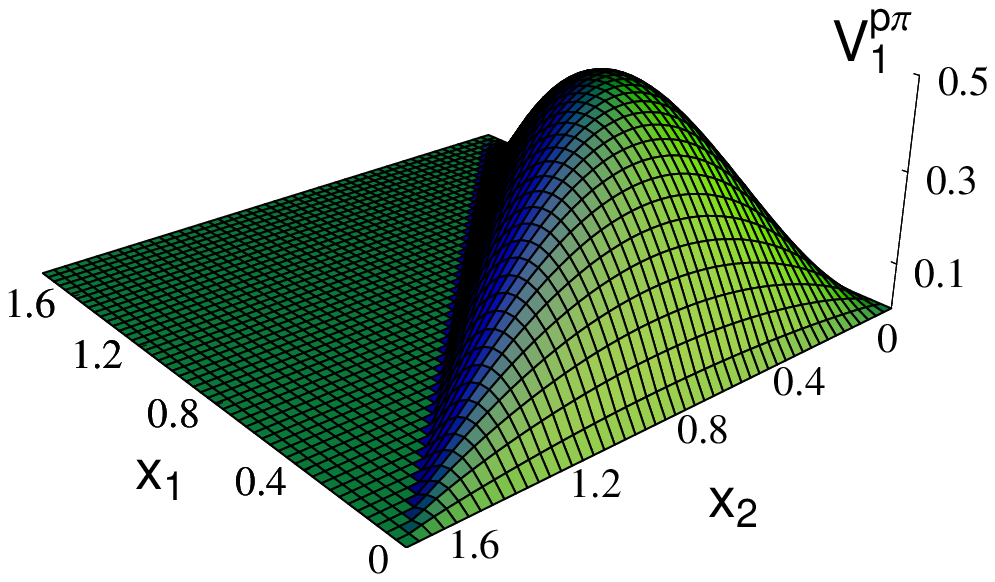,  width=18 pc}
\hspace{0.5 truecm}
\epsfig{file=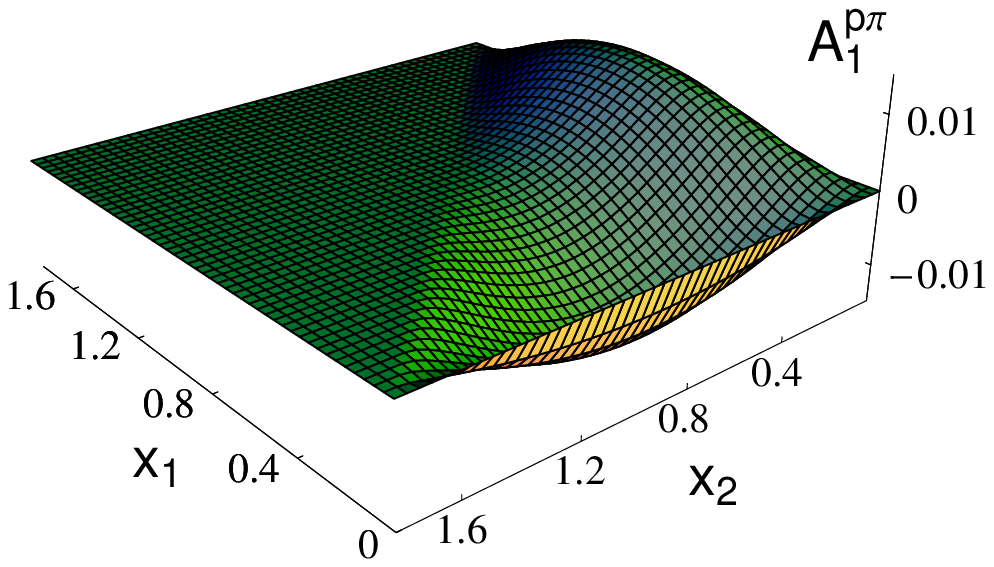,  width=18 pc}
\epsfig{file=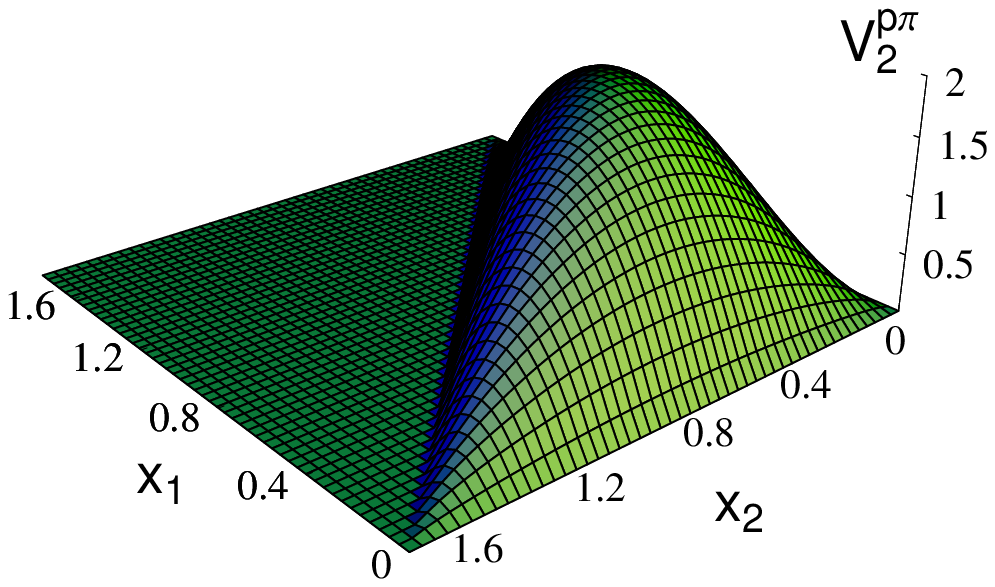,  width=18 pc}
\hspace{0.5 truecm}
\epsfig{file=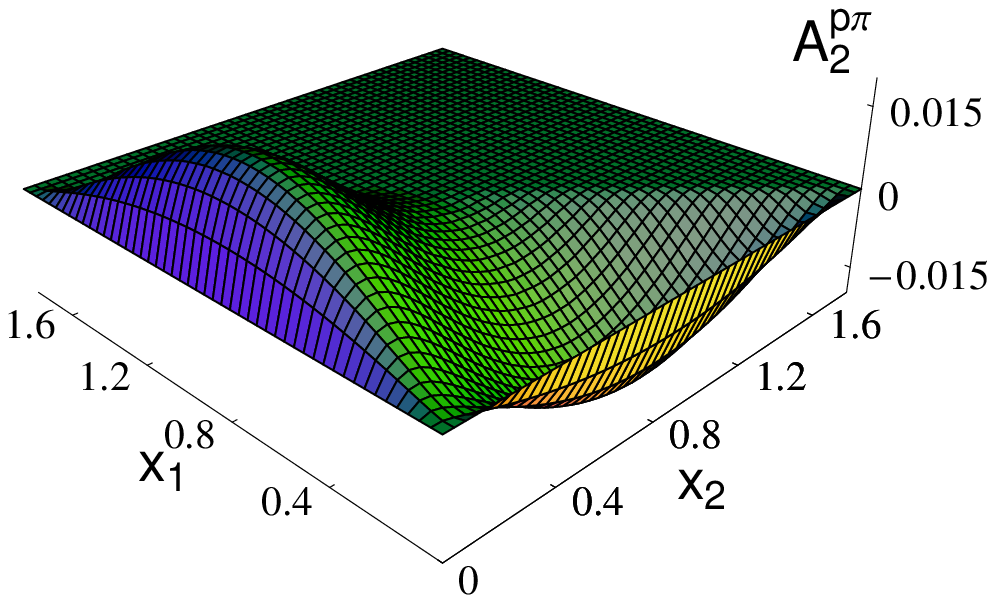,  width=18 pc}
	\caption{\label{Fig:V1-A1-V2-A2-p_pi}
The $p\rightarrow \pi^0$  transition distribution amplitudes  $V^{p\pi}_1$ (up left), $A^{p\pi}_1$ (up right), $V^{p\pi}_2$ (down left), $A^{p\pi}_2$ (down right) as function of 
$(x_1,x_2,2\xi-x_1-x_2)$ at fixed $\xi=0.9$ and $\Delta^2=-0.1$ GeV$^2$. }
\end{figure}

\begin{figure}[h!]
 	\centering
\epsfig{file=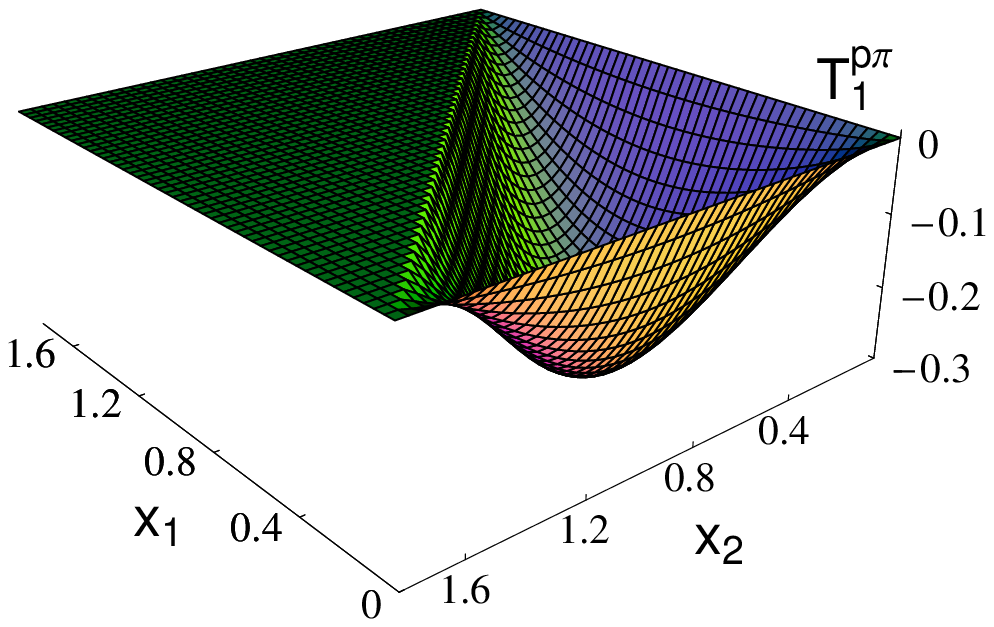,  width=18 pc}
\hspace{0.5 truecm}
\epsfig{file=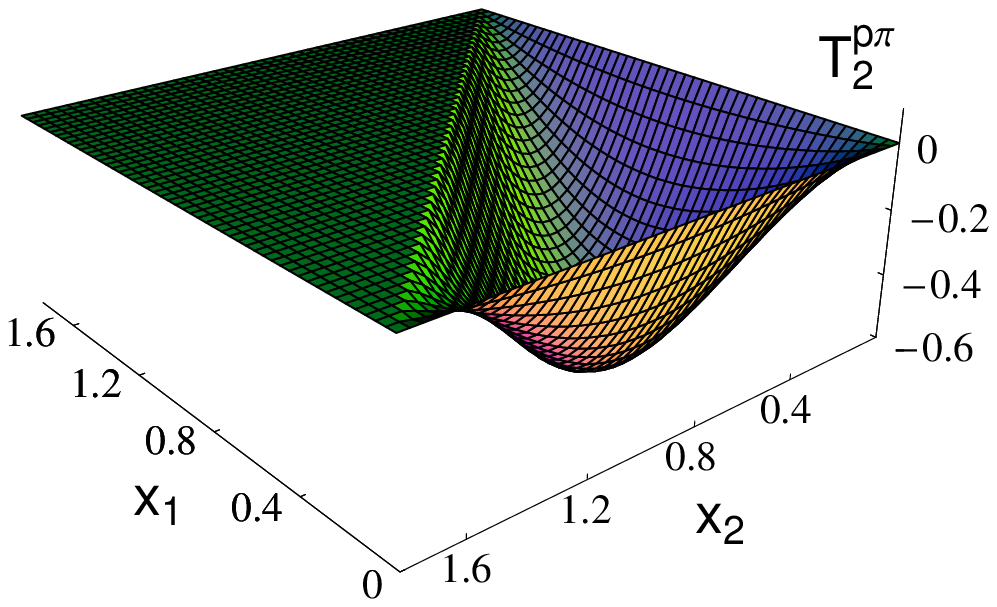,  width=18 pc}
\epsfig{file=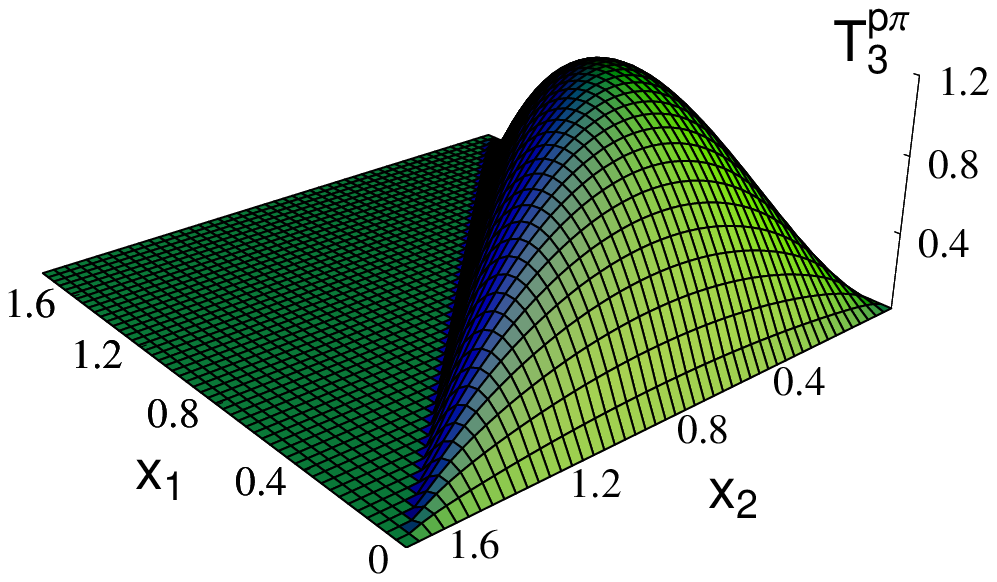,  width=18 pc}
\hspace{0.5 truecm}
\epsfig{file=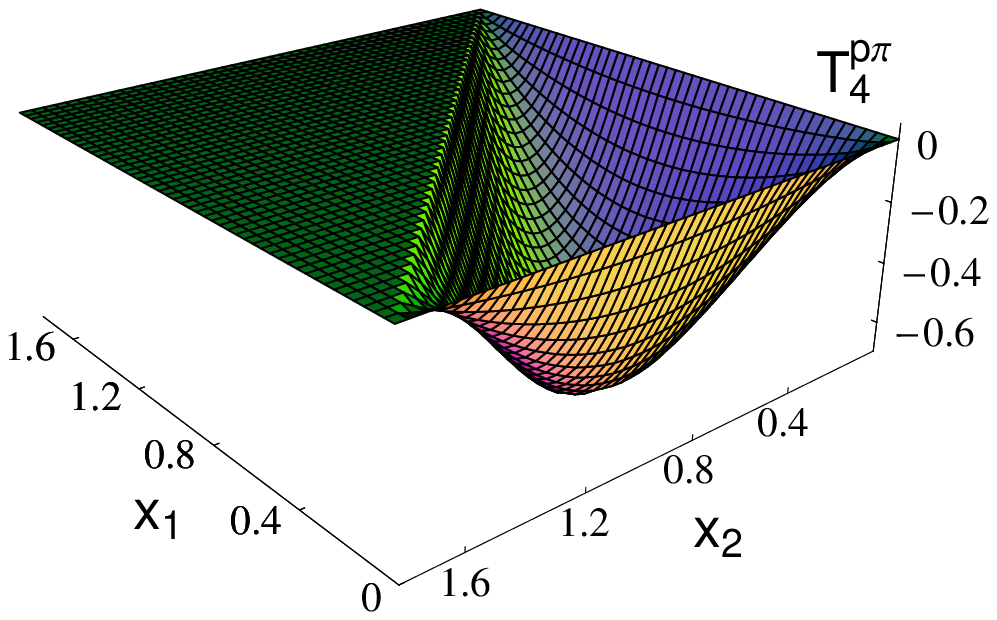,  width=18 pc}
	\caption{\label{Fig:T1-T2-T3-T4-p_pi}
The  $p\rightarrow \pi^0$ transition distribution amplitudes $T^{p\pi}_1$ (up left), $T^{p\pi}_2$ (up right), $T^{p\pi}_3$ (down left), and $T^{p\pi}_4$ (down left) as function of $(x_1,x_2,2\xi-x_1-x_2)$ at fixed $\xi=0.9$ and $\Delta^2=-0.1$ GeV$^2$.}
\end{figure}


\section{Concluding remarks}

In the light-cone description the nucleon state is decomposed in terms of $N$-parton Fock states with coefficients representing the momentum light-cone wave function of the $N$ partons. Since the constituent quark models work so well phenomenologically, in applications it is usually assumed that only the Fock components with a few partons have to be taken into account. One of such models has been studied in a series of papers to show that the parametrization of the LCWF up to five-parton components is already sufficient to account for the electroweak form factors~\cite{Pasquini:2007iz} and spin densities~\cite{Pasquini:2007xz} of the nucleon, as well as the observed asymmetries due to transverse momentum dependence of parton distributions~\cite{Pasquini:2008ax,Boffi:2009sh}, and to give a useful insight into the quark generalized parton distributions~\cite{Boffi:2002yy,Boffi:2003yj,Pasquini:2005dk,Pasquini:2006dv}.

As a further test of the model in this paper the nucleon distribution amplitudes and the nucleon-to-meson transition distribution amplitudes have been considered. At leading twist the nucleon DAs probe the three-quark content of the nucleon state with orbital angular momentum $L_z=0$ and the $N\to\pi$ TDAs probe the $q\bar q$ sea pair contribution responsible for the meson cloud surrounding the bare three-quark nucleon.

Assuming SU(6) symmetry the shape of the calculated nucleon DA is similar to the asymptotic DA, with a roughly symmetric contribution of the three quarks. This contrasts the results from QCD sum rules that push towards highly asymmetric quark contributions, but it is along the same lines of phenomenological models indicating that a less asymmetric DA is preferable to describe the nucleon form factors. Departures from the SU(6) symmetric model are necessary in the description of the neutron form factors~\cite{Pasquini:2007iz} and the large $x$ behavior of the neutron structure functions in deep inelastic processes~\cite{Boffi:2009sh}. Such SU(6) symmetry breaking contributions would lead to an asymmetric DA. This will be studied in a broader framework in a forthcoming paper~\cite{Pasquini:2008}. In any case, after evolution from the low-scale of the model to $Q^2=1$ GeV$^2$ the first and second DA moments, calculated within the present model and shown in Table~\ref{Moments_Evol}, already compare well with lattice QCD results~\cite{Lenz:2009ar}.

In contrast to the nucleon DAs that have been studied for a long time, only very recently attention to the nucleon-to-meson TDAs has been drawn, and the possibility of having some information from experiment has been suggested. Here, for the first time a model calculation has been presented for the eight leading twist $N\to\pi^0$ TDAs. They receive contribution from the fluctuations of the nucleon in ($p\pi^0$) and ($\Delta^+\pi^0$) subsystems and can be expressed as the convolution of the baryon DAs with the probability amplitude to find the corresponding baryon-meson component in the nucleon. The relative contribution of these components depends on the momentum transferred between the initial nucleon and the final pion as well as on the different spin configurations of the intermediate baryon. In particular, the $\Delta$ plays a special role in the case of the tensor TDAs which involve configurations with helicity $\pm3/2$, while the interplay of the nucleon and $\Delta$ contributions with helicity $\pm1/2$ determines the different shape of the vector and axial-vector TDAs.

Results have been shown under kinematic conditions reachable, e.g., at GSI-FAIR as proposed in Ref.~\cite{Lansberg:2007se}, but the model can easily and will be applied to other kinematics such as those proposed to study at Jlab~\cite{Lansberg:2007ec}.


\vspace{0.5cm}

\noindent{\bf Acknowledgements.}
We are grateful to J.P. Lansberg, L. Szymanowski and B. Pire for stimulating discussions and for the interest in this work.
The work  is part of the Research Infrastructure Integrating Activity ``Study of Strongly Interacting Matter'' (acronym HadronPhysics2, Grant Agreement n. 227431) under the Seventh Framework Programme of the European Community.




\appendix
\section{Spin components of the baryon light-cone wave functions}
\label{appendix}

In this Appendix we give the explicit results for the spin-dependent component
of the LCWFs of the proton and $\Delta$ state.

In the case of the proton, we have:

For the spin $\uparrow$ proton
\begin{eqnarray}
  \Xi^{p}_{\uparrow}\left(\uparrow,\downarrow,\uparrow\right)
&=&\frac{1}{\sqrt{6}}\prod_i\frac{1}{\sqrt{N(k_i)}}
\,(-a_1a_2a_3+k_1^L k_2^R a_3-2a_1k_2^Rk_3^L),
\label{first}
\\
  \Xi^{p}_\uparrow\left(\downarrow,\uparrow,\uparrow\right)
&=&\frac{1}{\sqrt{6}}\prod_i\frac{1}{\sqrt{N(k_i)}}
\,(-a_1a_2a_3+k_1^Rk_2^La_3 -2k_1^Ra_2k_3^L),
\\
  \Xi^{p}_\uparrow\left(\uparrow,\uparrow,\downarrow\right)
&=&\frac{1}{\sqrt{6}}\prod_i\frac{1}{\sqrt{N(k_i)}}
\,(2a_1a_2a_3+a_1 k_2^Lk_3^R+k_1^La_2k_3^R),
\\
  \Xi^{p}_\uparrow\left(\downarrow,\downarrow,\downarrow\right)
&=&\frac{1}{\sqrt{6}}\prod_i\frac{1}{\sqrt{N(k_i)}}
\,(-a_1 k^R_2 k_3^R -k^R_1 a_2 k_3^R+2 k_1^R k_2^R a_3)
\\
  \Xi^{p}_\uparrow\left(\uparrow,\downarrow,\downarrow\right)
&=&\frac{1}{\sqrt{6}}\prod_i\frac{1}{\sqrt{N(k_i)}}
\,(a_1 a_2k^R_3-k_1^L k_2^R k_3^R-2a_1k_2^Ra_3),
\\
  \Xi^{p}_\uparrow\left(\downarrow,\uparrow,\downarrow\right)
&=&\frac{1}{\sqrt{6}}\prod_i\frac{1}{\sqrt{N(k_i)}}
\,(-k_1^R k_2^Lk^R_3+a_1 a_2k_3^R-2k_1^Ra_2a_3),
\\
  \Xi^{p}_\uparrow\left(\uparrow,\uparrow,\uparrow\right)
&=&\frac{1}{\sqrt{6}}\prod_i\frac{1}{\sqrt{N(k_i)}}
\,(2a_1a_2k_3^L-a_1 k_2^L a_3-k_1^L a_2 a_3),
\\
  \Xi^{p}_\uparrow\left(\downarrow,\downarrow,\uparrow\right)
&=&\frac{1}{\sqrt{6}}\prod_i\frac{1}{\sqrt{N(k_i)}}
\,(k_1^R a_2 a_3+a_1 k_2^R a_3 +2 k_1^R k_2^R k_3^L);
\end{eqnarray}
For the spin $\downarrow$ proton
\begin{eqnarray}
  \Xi^{p}_\downarrow\left(\uparrow,\downarrow,\uparrow\right)
&=&\frac{1}{\sqrt{6}}\prod_i\frac{1}{\sqrt{N(k_i)}}
\,(a_1a_2 k_3^L- k_1^L k_2^R k_3^L -2 k_1^L a_2 a_3),
\\
  \Xi^{p}_\downarrow\left(\downarrow,\uparrow,\uparrow\right)
&=&\frac{1}{\sqrt{6}}\prod_i\frac{1}{\sqrt{N(k_i)}}
\,(a_1a_2 k_3^L- k_1^R k_2^L k_3^L -2 a_1 k_2^L a_3),
\\
  \Xi^{p}_\downarrow\left(\uparrow,\uparrow,\downarrow\right)
&=&\frac{1}{\sqrt{6}}\prod_i\frac{1}{\sqrt{N(k_i)}}
\,(k_1^L a_2a_3 +2 k_1^L k_2^L k_3^R + a_1 k_2^L a_3),
\\
  \Xi^{p}_\downarrow\left(\downarrow,\downarrow,\downarrow\right)
&=&\frac{1}{\sqrt{6}}\prod_i\frac{1}{\sqrt{N(k_i)}}
\,(-k_1^R a_2a_3 +2 a_1 a_2 k_3^R - a_1 k_2^R a_3),
\\
  \Xi^{p}_\downarrow\left(\uparrow,\downarrow,\downarrow\right)
&=&\frac{1}{\sqrt{6}}\prod_i\frac{1}{\sqrt{N(k_i)}}
\,(a_1 a_2 a_3-k_1^L k_2^R a_3+2k_1^L a_2 k_3^R),
\\
  \Xi^{p}_\downarrow\left(\downarrow,\uparrow,\downarrow\right)
&=&\frac{1}{\sqrt{6}}\prod_i\frac{1}{\sqrt{N(k_i)}}
\,(a_1a_2a_3 -k_1^R k_2^L a_3+2a_1 k_2^Lk_3^R),
\\
  \Xi^{p}_\downarrow\left(\uparrow,\uparrow,\uparrow\right)
&=&\frac{1}{\sqrt{6}}\prod_i\frac{1}{\sqrt{N(k_i)}}
\,(a_1 k_2^L k_3^L+k_1^L a_2 k_3^L-2 k_1^L k_2^L a_3),
\\
  \Xi^{p}_\downarrow\left(\downarrow,\downarrow,\uparrow\right)
&=&\frac{1}{\sqrt{6}}\prod_i\frac{1}{\sqrt{N(k_i)}}
\,(-2a_1a_2 a_3-a_1 k_2^R k_3^L - k_1^R a_2 k_3^L).
\end{eqnarray}

In the case of $\Delta$ we can have the following spin configurations:

For the $\Delta$ in the spin $\treh$ state, indicated with $\Uparrow$:
\begin{eqnarray}
  \Xi^{\Delta}_{\Uparrow}\left(\uparrow,\downarrow,\uparrow\right)
&=&\prod_i\frac{1}{\sqrt{N(k_i)}}
\,(-a_1k^R_2a_3),
\\
  \Xi^{\Delta}_\Uparrow\left(\downarrow,\uparrow,\uparrow\right)
&=&\prod_i\frac{1}{\sqrt{N(k_i)}}
\,(-k^R_1a_2a_3),
\\
  \Xi^{\Delta}_\Uparrow\left(\uparrow,\uparrow,\downarrow\right)
&=&
\prod_i\frac{1}{\sqrt{N(k_i)}}
\,(-a_1a_2k^R_3),
\\
  \Xi^{\Delta}_\Uparrow\left(\downarrow,\downarrow,\downarrow\right)
&=&
\prod_i\frac{1}{\sqrt{N(k_i)}}
\,(-k^R_1 k^R_2 k_3^R),
\\
  \Xi^{\Delta}_\Uparrow\left(\uparrow,\downarrow,\downarrow\right)
&=&
\prod_i\frac{1}{\sqrt{N(k_i)}}
\,(a_1 k^R_2k^R_3),
\\
  \Xi^{\Delta}_\Uparrow\left(\downarrow,\uparrow,\downarrow\right)
&=&
\prod_i\frac{1}{\sqrt{N(k_i)}}
\,(k_1^R a_2k^R_3),
\\
  \Xi^{\Delta}_\Uparrow\left(\uparrow,\uparrow,\uparrow\right)
&=&
\prod_i\frac{1}{\sqrt{N(k_i)}}
\,(a_1a_2a_3),
\\
  \Xi^{\Delta}_\Uparrow\left(\downarrow,\downarrow,\uparrow\right)
&=&
\prod_i\frac{1}{\sqrt{N(k_i)}}
\,(k_1^R k^R_2 a_3);
\end{eqnarray}
For the $\Delta$ in the spin $\uparrow$ state:
\begin{eqnarray}
  \Xi^{\Delta}_{\uparrow}\left(\uparrow,\downarrow,\uparrow\right)
&=&\frac{1}{\sqrt{3}}\prod_i\frac{1}{\sqrt{N(k_i)}}
\,(a_1a_2a_3-k_1^L k_2^Ra_3-a_1k_2^Rk_3^L),
\\
  \Xi^{\Delta}_\uparrow\left(\downarrow,\uparrow,\uparrow\right)
&=&\frac{1}{\sqrt{3}}\prod_i\frac{1}{\sqrt{N(k_i)}}
\,(a_1a_2a_3-k_1^Rk_2^La_3 -k_1^Ra_2k_3^L),
\\
  \Xi^{\Delta}_\uparrow\left(\uparrow,\uparrow,\downarrow\right)
&=&\frac{1}{\sqrt{3}}\prod_i\frac{1}{\sqrt{N(k_i)}}
\,(a_1a_2a_3-a_1 k_2^Lk_3^R-k_1^La_2k_3^R),
\\
  \Xi^{\Delta}_\uparrow\left(\downarrow,\downarrow,\downarrow\right)
&=&\frac{1}{\sqrt{3}}\prod_i\frac{1}{\sqrt{N(k_i)}}
\,(a_1 k^R_2 k_3^R +k^R_1 a_2 k_3^R+ k_1^R k_2^R a_3)
\\
  \Xi^{\Delta}_\uparrow\left(\uparrow,\downarrow,\downarrow\right)
&=&\frac{1}{\sqrt{3}}\prod_i\frac{1}{\sqrt{N(k_i)}}
\,(-a_1 a_2k^R_3+k_1^L k_2^R k_3^R-a_1k_2^Ra_3),
\\
  \Xi^{\Delta}_\uparrow\left(\downarrow,\uparrow,\downarrow\right)
&=&\frac{1}{\sqrt{3}}\prod_i\frac{1}{\sqrt{N(k_i)}}
\,(k_1^R k_2^Lk^R_3-a_1 a_2k_3^R-k_1^Ra_2a_3),
\\
  \Xi^{\Delta}_\uparrow\left(\uparrow,\uparrow,\uparrow\right)
&=&\frac{1}{\sqrt{3}}\prod_i\frac{1}{\sqrt{N(k_i)}}
\,(a_1a_2k_3^L+a_1 k_2^L a_3+k_1^L a_2 a_3),
\\
  \Xi^{\Delta}_\uparrow\left(\downarrow,\downarrow,\uparrow\right)
&=&\frac{1}{\sqrt{3}}\prod_i\frac{1}{\sqrt{N(k_i)}}
\,(-k_1^R a_2 a_3-a_1 k_2^R a_3 + k_1^R k_2^R k_3^L);
\end{eqnarray}
For the $\Delta$ in the spin $\downarrow$ state:
\begin{eqnarray}
  \Xi^{\Delta}_\downarrow\left(\uparrow,\downarrow,\uparrow\right)
&=&\frac{1}{\sqrt{3}}\prod_i\frac{1}{\sqrt{N(k_i)}}
\,(a_1a_2 k_3^L- k_1^L k_2^R k_3^L + k_1^L a_2 a_3),
\\
  \Xi^{\Delta}_\downarrow\left(\downarrow,\uparrow,\uparrow\right)
&=&\frac{1}{\sqrt{3}}\prod_i\frac{1}{\sqrt{N(k_i)}}
\,(a_1a_2 k_3^L- k_1^R k_2^L k_3^L + a_1 k_2^L a_3),
\\
  \Xi^{\Delta}_\downarrow\left(\uparrow,\uparrow,\downarrow\right)
&=&\frac{1}{\sqrt{3}}\prod_i\frac{1}{\sqrt{N(k_i)}}
\,(k_1^L a_2a_3 -k_1^L k_2^L k_3^R + a_1 k_2^L a_3),
\\
  \Xi^{\Delta}_\downarrow\left(\downarrow,\downarrow,\downarrow\right)
&=&\frac{1}{\sqrt{3}}\prod_i\frac{1}{\sqrt{N(k_i)}}
\,(-k_1^R a_2a_3 - a_1 a_2 k_3^R - a_1 k_2^R a_3),
\\
  \Xi^{\Delta}_\downarrow\left(\uparrow,\downarrow,\downarrow\right)
&=&\frac{1}{\sqrt{3}}\prod_i\frac{1}{\sqrt{N(k_i)}}
\,(a_1 a_2 a_3-k_1^L k_2^R a_3-k_1^L a_2 k_3^R),
\\
  \Xi^{\Delta}_\downarrow\left(\downarrow,\uparrow,\downarrow\right)
&=&\frac{1}{\sqrt{3}}\prod_i\frac{1}{\sqrt{N(k_i)}}
\,(a_1a_2a_3 -k_1^R k_2^L a_3-a_1 k_2^Lk_3^R),
\\
  \Xi^{\Delta}_\downarrow\left(\uparrow,\uparrow,\uparrow\right)
&=&\frac{1}{\sqrt{3}}\prod_i\frac{1}{\sqrt{N(k_i)}}
\,(a_1 k_2^L k_3^L+k_1^L a_2 k_3^L+ k_1^L k_2^L a_3),
\\
  \Xi^{\Delta}_\downarrow\left(\downarrow,\downarrow,\uparrow\right)
&=&\frac{1}{\sqrt{3}}\prod_i\frac{1}{\sqrt{N(k_i)}}
\,(a_1a_2 a_3-a_1 k_2^R k_3^L - k_1^R a_2 k_3^L);
\end{eqnarray}
For the $\Delta$ in the spin $-\treh$ state, indicated with $\Downarrow$:
\begin{eqnarray}
  \Xi^{\Delta}_\Downarrow\left(\uparrow,\downarrow,\uparrow\right)
&=&
\prod_i\frac{1}{\sqrt{N(k_i)}}
\,(k_1^L a_2 k_3^L),
\\
  \Xi^{\Delta}_\Downarrow\left(\downarrow,\uparrow,\uparrow\right)
&=&
\prod_i\frac{1}{\sqrt{N(k_i)}}
\,(a_1 k_2^L k^L_3),
\\
  \Xi^{\Delta}_\Downarrow\left(\uparrow,\uparrow,\downarrow\right)
&=&
\prod_i\frac{1}{\sqrt{N(k_i)}}
\,(k_1^L k_2^L a_3),
\\
  \Xi^{\Delta}_\Downarrow\left(\downarrow,\downarrow,\downarrow\right)
&=&\prod_i\frac{1}{\sqrt{N(k_i)}}
\,(a_1a_2a_3),
\\
  \Xi^{\Delta}_\Downarrow\left(\uparrow,\downarrow,\downarrow\right)
&=&
\prod_i\frac{1}{\sqrt{N(k_i)}}
\,(k_1^L a_2 a_3),
\\
  \Xi^{\Delta}_\Downarrow\left(\downarrow,\uparrow,\downarrow\right)
&=&\prod_i\frac{1}{\sqrt{N(k_i)}}
\,(a_1k_2^L a_3),
\\
  \Xi^{\Delta}_\Downarrow\left(\uparrow,\uparrow,\uparrow\right)
&=&\prod_i\frac{1}{\sqrt{N(k_i)}}
\,(k^L_1 k_2^L k_3^L),
\\
  \Xi^{\Delta}_\Downarrow\left(\downarrow,\downarrow,\uparrow\right)
&=&\prod_i\frac{1}{\sqrt{N(k_i)}}
\,(a_1a_2 k^L_3).\label{last}
\end{eqnarray}

Throughout Eqs.~(\ref{first})-(\ref{last}) we used the following 
definitions:
$a_i=(m+x_i M_0)$, $N(k_i)=[(m+x_i M_0)^2+k^2_{i\perp}]$,
$k_i^{R}=k_{i\,x}+ik_{i\,y}$, and  $k_i^{L}=k_{i\,x}-ik_{i\,y}$.

\clearpage



\end{document}